\documentclass[epjST]{svjour}
\usepackage{graphics}
\usepackage{graphicx}
\usepackage{dcolumn}
\usepackage{bm}

\usepackage{graphics}
\usepackage{cancel}
\usepackage{multirow}
\usepackage{longtable}
\usepackage{color}
\usepackage[normalem]{ulem} 
\usepackage{lipsum}
\usepackage{yfonts}

\def\be{\begin{equation}}
\def\ee{\end{equation}}
\newcommand{\bel}[1]{\begin{eqnarray}\label{#1}}
\newcommand{\eel}{\end{eqnarray}}
\def\barr{\begin{array}}
	\def\earr{\end{array}}
\def\beq{\begin{eqnarray}}
\def\eeq{\end{eqnarray}}
\def\bfig{\begin{figure}}
	\def\efig{\end{figure}}

\newcommand{\nn}{\nonumber}
\newcommand{\f}[2]{\frac{#1}{#2}}

\newcommand{\p}{\partial}

\newcommand{\tr}{{\rm tr}}
\newcommand{\rf}[1]{Eq.~(\ref{#1})}

\newcommand{\rfn}[1]{(\ref{#1})}

\def\LR{\left(} 
\def\RR{\right)}

\newcommand{\sh}[1]{\sinh#1}
\newcommand{\ch}[1]{\cosh#1}

\def\pmu{p^\mu}

\def\bmu{\beta_\mu}

\def\omnL{\omega_{\mu\nu}}
\def\omnU{\omega^{\mu\nu}}

\def\omnUD{\tilde {\omega}^{\mu\nu}}

\def\SmunuU{{\Sigma}^{\mu\nu}}

\def\S0iU{{\Sigma}^{0i}} 
 
\def\SmnU{{\Sigma}^{\mu\nu}}

\def\GLW{{\rm GLW}}

\def\nU{n_{(0)}}
\def\eU{\varepsilon_{(0)}}
\def\PU{P_{(0)}}
\def\sU{s_{(0)}}

\def\Cv{{\boldsymbol C}}

\usepackage{amssymb}

\usepackage{graphicx}
\usepackage{amsmath}
\usepackage{subfigure}
\usepackage{bbold}
\usepackage{yfonts}
\usepackage[colorlinks=true,linktocpage=true,linkcolor=blue,citecolor=blue]{hyperref}
\usepackage{placeins}
\usepackage{bm} 
\usepackage{nicefrac}
\usepackage{slashed}

\newcommand{\bea}{\begin{eqnarray}}
\newcommand{\eea}{\end{eqnarray}}

\def\LB{\left(}
\def\RB{\right)}
\def\LSB{\left[}
\def\RSB{\right]}

\newcommand{\pv}{{\boldsymbol p}}

\newcommand{\sv}{{\boldsymbol s}}

\def\LR{\left(} 
\def\RR{\right)}

\def\GLW{{\rm GLW}}

\def\CHI{\chi}
\def\nU{n_{(0)}}
\def\eU{\varepsilon_{(0)}}
\def\PU{P_{(0)}}
\def\sU{s_{(0)}}

\newcommand{\lab}[1]{\label{#1}}
\def\nn{\nonumber}

\def\pv{{\boldsymbol p}}

\def\omnL{\omega_{\mu\nu}}
\def\omnU{\omega^{\mu\nu}}

\def\omnUD{\tilde {\omega}^{\mu\nu}}

\def\epsUabgd{\epsilon^{\alpha \beta \gamma \delta}}

\def\TmnU{T^{\mu\nu}}

\def\SmnU{{\Sigma}^{\mu\nu}}

\def\S0iU{{\Sigma}^{0i}}

\def\n0{n_{(0)}}
\def\e0{\varepsilon_{(0)}}
\def\P0{P_{(0)}}

\def\TmnU{T^{\mu\nu}}                      

\def\fplusrsxp{f^+_{rs}(x,p)}

\def\fminusrsxp{f^-_{rs}(x,p)}

\def\ubarrp{{\bar u}_r(p)}

\def\usp{u_s(p)}
\def\urp{u_r(p)}

\def\vbarsp{{\bar v}_s(p)}

\def\vrp{v_r(p)}

\def\pv{{\boldsymbol p}}

\def\Cv{{\boldsymbol C}}

\def\Wxk{{\cal W}(x,k)}
\def\Weqxk{{\cal W}_{\rm eq}(x,k)}
\def\Weqpxk{{\cal W}^{+}_{\rm eq}(x,k)}
\def\Weqmxk{{\cal W}^{-}_{\rm eq}(x,k)}
\def\Weqpmxk{{\cal W}^{\pm}_{\rm eq}(x,k)}

\def\Fxk{{\cal F}(x,k)}

\def\Feqpmxk{{\cal F}^{\pm}_{\rm eq}(x,k)}

\def\Pxk{{\cal P}(x,k)}

\def\Peqpmxk{{\cal P}^{\pm}_{\rm eq}(x,k)}

\renewcommand\sout{\bgroup\markoverwith{\textcolor{blue}{\rule[0.5ex]{2pt}{0.4pt}}}\ULon}

\DeclareMathOperator{\sech}{sech}

\begin{document}
\title{New Developments in Relativistic Fluid Dynamics with spin}
\author{Samapan Bhadury\inst{1}\fnmsep\thanks{\email{samapan.bhadury@niser.ac.in}} 
\and Jitesh Bhatt\inst{2}\fnmsep\thanks{\email{jeet@prl.res.in}}
\and Amaresh Jaiswal\inst{1}\fnmsep\thanks{\email{a.jaiswal@niser.ac.in}} \and Avdhesh Kumar \inst{1}\fnmsep\thanks{\email{avdhesh.kumar@niser.ac.in}}}
\institute{School of Physical Sciences, National Institute of Science Education and Research, HBNI, Jatni-752050, \and Physical Research Laboratory, Navrangpura, Ahmedabad 380 009, India}
\abstract{
 In this work, we briefly review the progress made in the formulation of hydrodynamics with spin with emphasis on the application to the relativistic heavy-ion collisions. In particular,  we discuss the formulation of hydrodynamics with spin for perfect-fluid and the first order viscous corrections with some discussion on the calculation of spin kinetic coefficients.  Finally, we apply relativistic hydrodynamics with spin to the relativistic heavy-ion collisions to calculate the spin polarization of $\Lambda$-particles.
} 
\maketitle
\section{Introduction}
\label{intro}
In ultra-relativistic non-central heavy-ion collisions colliding nuclei carry a
huge orbital angular momentum. 
Soon after the collision, a substantial portion of
this orbital angular momentum gets deposited in the interaction zone which can further be transformed from initial purely orbital to the spin form. The latter can be displayed in the spin polarization of the emerging particles. Indeed, experimental results show that the spin of various emitted particles  ($\Lambda$, $K^*$, $\phi$ etc) produced during the collision aligned with the global angular momentum direction~\cite{STAR:2017ckg,Adam:2018ivw,Acharya:2019vpe}. 
Theoretically, first predictions of global polarization of produced $\Lambda$ hyperons, based on spin-orbit interaction and perturbative-QCD inspired model were reported in Refs.~\cite{Voloshin:2004ha} and \cite{Liang:2004ph,Liang:2004xn,Betz:2007kg}.  
In these works, a significant polarization effect of the order of 10\% was reported.
Later, based on relativistic 
hydrodynamics and assuming local thermodynamic equilibrium of the spin degrees of freedom, a smaller polarization of about 1\% \cite{Becattini:2007sr,Becattini:2013vja,Becattini:2013fla,Becattini:2007nd,Becattini:2016gvu,Becattini:2015ska,Karpenko:2016jyx,Xie:2017upb,Pang:2016igs,Becattini:2017gcx} was predicted which was later confirmed by STAR~\cite{STAR:2017ckg,Adam:2018ivw}.
However, unfortunately, the same models~\cite{Becattini:2017gcx,Becattini:2020ngo} were unable to describe the experimentally measured longitudinal polarization of  $\Lambda$ particles~\cite{Adam:2018ivw,Niida:2018hfw}. It was seen the oscillations of the longitudinal polarization of $\Lambda$-hyperons as a function of the azimuthal angle as observed by the STAR experiment~\cite{Niida:2018hfw} has an opposite sign with respect to the results obtained using relativistic hydrodynamic models with thermalized spin degree of freedom.

So far, the global spin polarization of $\Lambda$ and $\bar\Lambda$-hyperons described by relativistic hydrodynamics (perfect or dissipative) make use of the fact that spin polarization effects are governed by thermal vorticity, $\varpi_{\mu \nu}= -\frac{1}{2} (\partial_\mu \beta_\nu-\partial_\nu \beta_\mu)$, where four vector $\beta_\mu$ is  defined by the ratio of the flow four-vector $u_\mu$ to the local temperature $T$, {\it i.e.} $\beta_\mu = u_\mu/T$~\cite{Becattini:2016gvu,Karpenko:2016jyx,Becattini:2017gcx,Weickgenannt:2020aaf}. However, from the general thermodynamics it is expected that spin polarization effects, governed by the tensor $\omega_{\mu \nu} $ (namely spin polarization tensor~\cite{Becattini:2018duy}), that can be independent of the thermal vorticity. 
This indicates the need for a new hydrodynamic approach that allows for the spin polarization tensor to be considered as an independent hydrodynamical variable. This new approach is referred as the hydrodynamics of spin polarized fluids or spin-hydrodynamics. Initial steps in the direction to formulate the perfect-fluid versions of hydrodynamics of spin polarized fluids have already been made  in a series of Refs.~\cite{Weickgenannt:2020aaf,Becattini:2018duy,Florkowski:2017ruc,Florkowski:2017dyn} (for applications see also Refs. \cite{Singh:2020rht,Singh:2021man,Jaiswal:2020hvk}).  Very recently, some progress has also been made where the dissipative effects are explicitly considered by introducing the collisions of particles. Refs.~\cite{Weickgenannt:2020aaf,Speranza:2020ilk,Shi:2020htn}.

  Any formulation of hydrodynamics that incorporates spin degree of freedom has to deal with including some quantum features. It should be noted here in the non-relativistic physics quantum hydrodynamics has been studied extensively; see Ref.~\cite{Haas:2011} and references cited therein. There exist several ways, perhaps equivalent, to describe the spin relativistic hydrodynamics: 1) Covariant techniques based on first deriving a Boltzmann equation from the quantum field theory. Then hydrodynamics is obtained by taking various moments of the kinetic equation \cite{Denicol:2012cn,Jaiswal:2013npa,Jaiswal:2013vta,Dash:2017rhg,Mohanty:2018eja,Dash:2020vxk}. 
  2) Approach based on Wigner function where one starts from spin-1/2 particles by constructing the kinetic model from Dirac equations \cite{Kharzeev:2007jp,Fukushima:2008xe,Kharzeev:2013ffa,Hirono:2014oda,Li:2014bha,Kharzeev:2015znc,Li:2016tel,Liu:2019krs,Gao:2020vbh,Liu:2020ymh,Li:2020vwh,Hattori:2019ahi,Yang:2020hri}.
  3) Lagrangian effective field theory techniques \cite{Montenegro:2017lvf,Montenegro:2017rbu,Montenegro:2018bcf,Montenegro:2019tku,Gallegos:2020otk,Gallegos:2021bzp}. 4) Approach based on the general thermodynamics where one derive equations of relativistic hydrodynamic with spin on the basis of an entropy-current analysis~\cite{Hattori:2019lfp}.  
  5) One can also derive the relativistic fluid equations Directly from the Dirac equation \cite{Asenjo:2011,Takabayasi:1957}. 
  In this approach one can write ``fluidization" of  Dirac equation by writing observables using several bilinear covariant. Next, the macroscopic fluid variables are constructed using ensemble average  over N-particle states. This procedure is somewhat complicated but it produces the correct non-relativistic limits of the spin-hydrodynamics. This work has recently been applied to some astrophysical scenario where the parity violating neutrino-electron interaction giving spin-dependent hydrodynamics was used to understand the pulsar kicks \cite{Bhatt:2016irk}.
 
 The prime focus of this review is to discuss the progress made in formulation of the framework of hydrodynamics for spin polarized fluids and its applications to heavy ion collisions. We organize this review paper  as follows: First, we briefly review Wigner function approach to formulate perfect-fluid hydrodynamics with spin in section \ref{sec:2}. In section \ref{sec:3}, we show such a frame work can also be derived using the classical treatment of spin degrees of freedom while in section \ref{sec:4} we extend the classical approach to include dissipation. In section \ref{sec:5}, we discuss applications of hydrodynamics with spin to heavy ion collisions. In section \ref{sec:6} we give a brief summary. 
\section{Wigner function approach to formulate perfect fluid hydrodynamic with spin}
\label{sec:2}  
Concept of Wigner function and its semiclassical expansion has been successfully used in past to construct the classical limit of quantum kinetic equations \cite{Elze:1986qd,Vasak:1987um,elze1989quark,Florkowski:1995ei,Zhuang:1995pd,Alexandrov:2020zsj}. Finding a general analytical solution of the full quantum kinetic equations using the Wigner function appears to be a highly non-trivial task. However the semiclassical expansion reduces this difficulty of an otherwise complicated theory as shown in a series of papers in the Refs.~\cite{Sheng:2017lfu,Sheng:2018jwf,Sheng:2020oqs,Sheng:2019ujr,Weickgenannt:2019dks} and still provide important physical insights. In this section we briefly review recently introduced equilibrium Wigner functions for particles with spin-1/2 that are used in the quantum kinetic equations.  Subsequently, in local thermodynamic equilibrium, we discuss, a simple procedure to formulate hydrodynamic framework for spin polarized fluids based on the semiclassical expansion of Wigner functions. 
\subsection{Equilibrium Wigner functions for particles with spin-1/2}
\label{subsec:2.1}
We start with the phase-space
distribution functions $f^{\pm}_{rs}(x,p)$ for particles ($+$) and antiparticles ($-$) with spin 1/2 at local thermodynamical equilibrium as introduced by Becattini et al. in Ref.~\cite{Becattini:2013fla}. These $f^{\pm}_{rs}(x,p)$ are the generalization of scaler single particle equilibrium distribution function 
in terms of $2\times2$ hermitian matrices in the spin space at each value of the space-time position $x$ and momentum four-vector $p$. 
\beq
\fplusrsxp \!\!=\!\!
\frac{1}{2m} \ubarrp X^+ \usp, ~ \fminusrsxp\!\!=\!\!- \frac{1}{2m}\vbarsp X^- \vrp .\nn 
\eeq	
In the above expressions,  $\urp $ and $\vrp$ are Dirac bispinors,  $r$ and $s$ are the spin indices running from 1 to 2, $m$ is the (anti-)particle mass and the objects $X^{\pm}$ are $4\times4$ matrices given by the formula 
\beq
X^{\pm} =  \exp\left[\pm \xi(x) - \bmu(x) \pmu \pm \f{1}{2} \omega_{\mu\nu}\Sigma^{\mu\nu}\right] \nn
\eeq
where $\xi=\frac{\mu}{T}$, is the ratio of the chemical potential $\mu$ to  temperature $T$ while $\beta^{\nu}=u^{\mu}/T$ with $u^{\mu}$ being the  flow four vector. The quantity, $\omnL$ is known as spin polarization tensor while $\SmunuU = (i/4) [\gamma^\mu,\gamma^\nu]$ as the Dirac spin operator. 

As discussed in Refs.~\cite{Florkowski:2017ruc,Florkowski:2017dyn}, if we assume that the spin polarization tensor $\omnL$ fulfills the conditions, $\omnL \omnU \geq 0$ and $\omnL \omnUD = 0$, where $\omnUD=\frac{1}{2}\epsilon^{\mu\nu\alpha\beta}\omega_{\alpha\beta}$ is the dual of $\omnU$, we can introduce a new parameter $\zeta  = \f{1}{2} \sqrt{ \frac{1}{2} \omnL \omnU }$. The parameter $\zeta$ can be interpreted as the ratio of spin  potential to temperature~\cite{Florkowski:2017ruc}.
The functions $f^{\pm}_{rs}(x,p)$ can be utilized to determine the corresponding equilibrium Wigner functions $\Weqpmxk$. Using the formula derived in Ref ~\cite{DeGroot:1980dk} the equibirium Wigner are given by
%
\beq
\Weqpxk &=& \frac{1}{2} \sum_{r,s=1}^2 \int dP\,
\delta^{(4)}(k-p) u^r(p) {\bar u}^s(p) f^+_{rs}(x,p),\label{eq:wig+} \\
\Weqmxk &=&-\frac{1}{2} \sum_{r,s=1}^2 \int dP\,
\delta^{(4)}(k+p) v^s(p) {\bar v}^r(p) f^-_{rs}(x,p),\label{eq:wig-}
\eeq	
where $dP = \frac{d^3p}{(2 \pi )^3 E_p}$ is the Lorentz invariant integration measure in mometum space with $E_p = \sqrt{m^2 + \pv^2}$ as the on-mass-shell particle energy. Argument, $x$ is the space-time coordinate and $k^\mu = (k^0, \mathbf{k})$ is the four momentum which is not necessarily on the mass shell.

Wigner functions given by above Eqs. (\ref{eq:wig+}) and (\ref{eq:wig-}) are the $4\times4$ matrices which satisfy the relation $\Weqpmxk =\gamma_0 \Weqpmxk^\dagger \gamma_0$. Therefore, they can always be expressed
with the help of 16 independent generators of the Clifford algebra~\cite{Elze:1986qd,Vasak:1987um}
%
\beq
\Weqpmxk = \f{1}{4} \Big[&& \Feqpmxk + i \gamma_5 \Peqpmxk + \gamma^\mu {\cal V}^\pm_{{\rm eq}, \mu}(x,k) \nn\\
&&+ \gamma_5 \gamma^\mu {\cal A}^\pm_{{\rm eq}, \mu}(x,k)
+ \SmnU {\cal S}^\pm_{{\rm eq},\mu \nu}(x,k) \Big]. \label{eq:equiwfn}
\eeq
Various coefficient functions $\Feqpmxk$, $\Peqpmxk$, ${\cal V}^\pm_{{\rm eq}, \mu}(x,k)$, ${\cal A}^\pm_{{\rm eq}, \mu}(x,k)$, ${\cal S}^\pm_{{\rm eq},\mu \nu}(x,k) $ appearing in the above expression, are known as scalar, pseudo-scalar, vector, axial-vector and tensor components of Wigner function, can be obtained by contracting $\Weqpmxk$ with appropriate gamma matrices and then taking the trace~\cite{Weickgenannt:2019dks,Florkowski:2018ahw}. The total Wigner function is given by the sum of the particle and antiparticle contributions {\it i.e.} $\Weqxk = \Weqpxk + \Weqmxk$. 
\subsection{$\hbar$-expansion} 
\label{subsec:2.2}
A decomposition similar to \rf{eq:equiwfn} can naturally be used for any arbitrary Wigner function $\Wxk$. Thus, we can write 
 \beq
\Wxk&=&\f{1}{4} \Big[ \Fxk + i \gamma_5 \Pxk + \gamma^\mu {\cal V}_{\mu}(x,k) \nn\\
&&+ \gamma_5 \gamma^\mu {\cal A}_{\mu}(x,k)
+ \SmnU {\cal S}_{\mu \nu}(x,k) \Big]. \label{eq:equiwfn1}
\eeq
For the case when there are no mean fields, $\Wxk$ satisfies the following equation \cite{Vasak:1987um}
\bel{eq:eqforW}
\left(\gamma_\mu K^\mu - m \right) {\cal W}(x,k) = C[{\cal W}(x,k)]; \qquad K^\mu = k^\mu + \frac{i \hbar}{2} \,\p^\mu,
\eel
where $ C[{\cal W}(x,k)]$ on the right hand side represents the collision term. In case of the global or local equilibrium the collision term vanishes. In this case, solution of Eq.~(\ref{eq:eqforW}) can be written in a series of $\hbar$,
\beq
{\cal X} = {\cal X}^{(0)}  + \hbar {\cal X}^{(1)}  +  \hbar^2 {\cal X}^{(2)}   + \cdots;   {\cal X} \in \{{\cal F}, {\cal P}, {\cal V}_\mu,{\cal A}_\mu,  {\cal S}_{\nu\mu} \}\label{eq:semi}
\eeq
Using Eqs. (\ref{eq:equiwfn1}), (\ref{eq:eqforW}), (\ref{eq:semi}) and keeping the terms upto first order in $\hbar$ expansion we can get the following kinetic equations for the coefficient functions ${\cal F}_{(0)}(x,k)$ and ${\cal A}^\nu_{(0)} (x,k)$~\cite{Florkowski:2018ahw}
\bel{eq:kineqs}
k^\mu \p_\mu {\cal F}_{(0)}(x,k) = 0,~ k^\mu \p_\mu \, {\cal A}^\nu_{(0)} (x,k) = 0, ~
k_\nu \,{\cal A}^\nu_{(0)} (x,k) =0.\nn
\eel
Here we note that functions ${\cal F}^{(0)}$ and ${\cal A}^{(0)}_\mu$ are basic independent ones. Kinetic equations for other coefficient functions can be easily derived using these two. Moreover, the algebraic structures of zeroth-order equations obtained from the semi-classical expansion of the Wigner function are similar to equations of the equilibrium coefficient functions. Therefore, we can assume ${\cal X}^{(0)}$ by ${\cal X}_{\rm eq}$. In this way following Boltzmann-like kinetic equations for the equilibrium coefficient functions ${\cal F}_{\rm eq}$ and ${\cal A}^\nu_{\rm eq}$ can be obtained
\bel{eq:kineqFC1}
k^\mu \p_\mu {\cal F}_{\rm eq}(x,k) = 0, \qquad k^\mu \p_\mu \, {\cal A}^\nu_{\rm eq} (x,k) = 0, \qquad k_\nu \,{\cal A}^\nu_{\rm eq}(x,k) = 0.\nn
\eel
Using the expressions of ${\cal F}_{\rm eq}$ and ${\cal A}^\nu_{\rm eq}$~\cite{Florkowski:2018ahw} one can see that these equations are exactly fulfilled if $\partial_{\mu}\beta_{\nu}-\partial_{\nu}\beta_{\mu}=0$ while parameters $\xi$ and $\omega_{\mu \nu}$ are constant. The equation for $\beta_{\mu}$ field is well known Killing equation which have the solution of the form  $\beta_{\mu}=b^{0}_{\mu}+\varpi_{\mu\nu}x^{\nu}$ with both $b^{0}_{\mu}$ and thermal vorticity $\varpi_{\mu\nu}$ being constant. Thus, we can conclude that in case of global equilibrium both spin polarization tensor $\omega_{\mu \nu}$ and  thermal vorticity $\varpi_{\mu \nu}$ are constant, however, nothing can be said about whether the two are equal. Here, we would like to emphasize the fact that in presence of the mean-fields and collisions spin polarization can be exactly equal to thermal vorticity as shown in Refs.~\cite{Weickgenannt:2020aaf,Weickgenannt:2019dks}   

\subsection{Formulation of perfect-fluid relativistic hydrodynamics with spin} 
\label{subsec:2.3}
Perfect-fluid hydrodynamics is govern by equations representing conservation law in local thermodynamic equilibrium. For a system with particles and anti-particles (without spin), the conserved quantities are the energy-momentum tensor ($T^{\mu\nu}$) and charge current ($N^\mu$). However, while considering particles with spin, one has to consider an additional conserved quantity: the spin tensor ($S^{\lambda, \mu\nu}$)~\cite{Florkowski:2018ahw,Florkowski:2018fap} which is the result of total angular momentum conservation. For a recent review on the subject see Refs.~ \cite{Becattini:2020ngo,Speranza:2020ilk,Florkowski:2018fap}. In following we shall obtain these quantities one by one and show that these are conserved.
\subsubsection{Charge current} 
\label{subsec:2.3.1}
The charge current ${\cal N}^{\alpha}(x)$ is related to Wigner function given by the following formula~\cite{DeGroot:1980dk}
\beq
{\cal N}^{\alpha}=\tr \int d^4k \, \gamma^{\alpha}\Wxk=\int d^4k \,{\cal V}^{\alpha} 
\eeq
In local equilibrium, we keep the terms upto $\hbar$ is ${\cal V}^{\alpha}$ and find, ${\cal N}^{\alpha}_{\rm eq}={N}^{\alpha}_{\rm eq}+\delta {N}^{\alpha}_{\rm eq}$ with  $\delta {N}^{\alpha}_{\rm eq}$ is the first order in $\hbar$ correction in the charge current. Note that  $\partial_{\alpha}\delta {N}^{\alpha}_{\rm eq}=0$. Thus the conserved charge current is given by $\partial_{\alpha}{N}^{\alpha}_{\rm eq}=0$ where,
\beq
{\cal N}^{\alpha}_{\rm eq}&=&\int d^4k \,{\cal V}^{\alpha}_{\rm eq} = 4 \cosh(\zeta) \, \sinh(\xi)\, \int dP \, p^{\alpha} \,  e^{- \beta \cdot p}
 \label{eq:neqs}
\eeq
Carrying out the integration over momentum charge current can be written as 
\beq
N^\alpha_{\rm eq} = n u^\alpha, \label{eq:cc}
\eeq
where 
\beq
n = 4 \cosh(\zeta) \, \sinh(\xi)\, \n0(T) \label{eq:nden}
\eeq
is the net charge density~\cite{Florkowski:2017ruc} while the quantity $\n0(T)$ is number density of spinless, neutral massive Boltzmann particles. It is defined in terms of the thermal average
\beq
\n0(T)= \langle(u\cdot p)\rangle_0 ,\label{eq:N0}
\eeq
where
\bel{avdef}
\langle \cdots \rangle_0 \equiv \int dP  (\cdots) \,  e^{- \beta \cdot p}.
\eel
Evaluating Eq.~(\ref{eq:N0}) we get
\bea
n_{(0)}(T) &=& \int dP  (u\cdot p) \,  e^{- \beta \cdot p} =I_{{10}}^{(0)} \nn \\
&=& \frac{1}{2 \pi ^2} \hat{m}^2 T^3 K_2 (\hat{m}) \,,
\eea
where $\hat{m}=m/T$ and thermodynamic integrals $I_{nq}^{(r)}$ defined in Appendix \ref{sec:thermint}.

\subsubsection{Energy-mometum tensor} 
\label{subsec:2.3.2}
The expression for energy-momentum in the GLW formulation are given by~\cite{DeGroot:1980dk}
\beq
T^{\mu\nu}_{\rm GLW}&=&\frac{1}{m}\tr\int d^4k \, k^{\mu}k^{\nu}\Wxk=\frac{1}{m}\int d^4k \, k^{\mu}k^{\nu}\Fxk
\eeq
Keeping the above equation upto first order in $\hbar$ and replacing ${\cal F}_{(0)}$ by ${\cal F}_{\rm eq}$  we can obtain
\beq
T^{\mu\nu}_{\rm GLW}&=&4 \cosh(\zeta) \, \cosh(\xi)\, \int dP \, p^{\mu} p^{\nu} \,  e^{- \beta \cdot p}
\eeq
After carrying out the momentum integration, $T^{\mu\nu}_{\rm GLW}$ can be expressed as
\beq
T^{\mu\nu}_{\rm GLW}(x) &=& \varepsilon u^\mu u^\nu - P \Delta^{\mu\nu},\label{Eq:Tmunu}
\eeq
where $\varepsilon$ and $P$ are the net energy density and pressure. They are expressed as follows
\bel{enden1}
\varepsilon = 4 \,\cosh(\xi) \, \e0(T)
\eel
and
\bel{prs1}
P = 4 \, \cosh(\xi) \, \P0(T),
\eel
The auxiliary quantities $\e0(T)$ and $\P0(T)$ are the energy density and pressure of the spinless, neutral massive Boltzmann particles which are defined by the following thermal average
\bel{enden0}
\e0(T) &=& \langle(u\cdot p)^2\rangle_0 
\eel
and
\bel{prs0}
\P0(T) = -(1/3) \langle \left[ p\cdot p - (u\cdot p)^2 \right] \rangle_0. 
\eel
In Eq.~(\ref{Eq:Tmunu}), second rank tensor object $\Delta^{\alpha\beta}$ is a projection operator which is orthogonal to the fluid four velocity $u^{\mu}$. 

Thermal average in Eqs. (\ref{enden0}), (\ref{prs0}) can be easily computed with help of thermodynamic integrals $I_{nq}^{(r)}$ as defined in Appendix \ref{sec:thermint}
\bea
\varepsilon_0(T) &=& \int{dP}\,(u\cdot p)^2 e^{- \beta \cdot p}=I^{(0)}_{20} \nn \\
&=& \frac{1}{2 \pi ^2}{\hat{m}^2 T^4 \left(3 K_2 (\hat{m})+\hat{m}K_1 (\hat{m}\right)}
\eea
and
\bea
P_0(T) &=& -\frac{1}{3}\Delta_{\mu\nu} \int{dP}\, p^{\mu}p^{\nu} e^{- \beta \cdot p} \nn \\ 
&=& -\frac{1}{3}\int{dP}\,( p\cdot p-(u\cdot p)^2)e^{- \beta \cdot p}=-I^{(0)}_{21} \nn \\
&=& \frac{1}{2 \pi ^2}{\hat{m}^2 T^4 K_2 (\hat{m})}=n_0 T.
\eea
\subsubsection{Spin tensor}
In the GLW formulation, spin tensor is related to Wigner function by following expression
\beq
S^{\lambda,\mu\nu}_{\rm GLW}&=&\frac{\hbar}{4}\int {d^k}\tr\Bigg[\left(2\{\Sigma^{\mu\nu},\gamma^{\lambda}\}+\frac{2i}{m}\left(\gamma^{[\mu}k^{\nu]}\gamma^{\lambda}-\gamma^{\lambda}\gamma^{[\mu}k^{\nu]}\right)\right)\Wxk\Bigg]
\eeq
Note that above equation is already in first order in $\hbar$. Therefore, in the  local equilibrium we can take leading order expression for Wigner function and replace it by $\Weqxk$. After performing trace and carrying out momentum integration we obtain
\bea
S^{\lambda , \mu \nu }_{\rm GLW}
&=&  {\cal C} \left( n_{(0)}(T) u^\lambda \omega^{\mu\nu}  +  S^{\lambda , \mu \nu }_{\Delta\GLW} \right).
\label{eq:Smunulambda_de_Groot2}
\eea 
where ${\cal C}=\hbar\frac{\sinh{\zeta}}{\zeta} \ch{\xi}$, while the auxiliary tensor $S^{\lambda , \mu \nu }_{\Delta\GLW}$ is expressed by 
\beq
S^{\alpha, \beta\gamma}_{\Delta\GLW} 
=  {\cal A}_{(0)} \, u^\alpha u^\delta u^{[\beta} \omega^{\gamma]}_{~\delta}
+{\cal B}_{(0)} \, \Big( 
u^{[\beta} \Delta^{\alpha\delta} \omega^{\gamma]}_{~\delta}
+ u^\alpha \Delta^{\delta[\beta} \omega^{\gamma]}_{~\delta}
+ u^\delta \Delta^{\alpha[\beta} \omega^{\gamma]}_{~\delta}\Big),\label{SDeltaGLW} 
\eeq
where
\beq 
{\cal B}_{(0)} &=&-\frac{2}{\hat{m}^2}  \frac{\varepsilon_{(0)}(T)+P_{(0)}(T)}{T}=-\frac{2}{\hat{m}^2} s_{(0)}(T)\label{coefB}
\eeq
and
\beq
{\cal A}_{(0)} &=&\frac{6}{\hat{m}^2} s_{(0)}(T) +2 n_{(0)} (T) = -3{\cal B}_{(0)} +2 n_{(0)}(T),\nn\\
\label{coefA}
\eeq
In Eqs. (\ref{coefB}) and (\ref{coefA}) quantity, $\sU =\LR\eU+\PU\RR / T$ is the entropy density of spin-0, neutral massive Boltzmann particles. 
Here we note that since GLW version of energy-momentum tensor as defined above is symmetric, the GLW spin tensor should be separately conserved {\it i.e.} $\partial_{\lambda}S^{\lambda,\mu\nu}_{\rm GLW}=0$. 
Now the conservation law for the charge current, energy momentum tenosr and spin tensor defined can be obtained by taking certain moments (as defined below) of kinetic equations in case of local thermodynamic equilibrium Ref.~\cite{Florkowski:2018ahw}.  
\bel{eq:kineqFt}
\int d^4k k^\mu \p_\mu {\cal F}_{\rm eq}(x,k) = 0 \qquad \Rightarrow \p_\mu N^\mu_{\rm GLW}(x) = 0.  
\eel
	\bel{eq:kineqAt}
	\int d^4k k^\mu k^\nu \p_\mu {\cal F}_{\rm eq}(x,k) = 0 \qquad \Rightarrow \p_\mu T^{\mu\nu}_{\rm GLW}(x) = 0.  
	\eel
	\bel{eq:kineqSC1}
&& \int d^4k \epsilon^{\mu\gamma\delta\eta} k_{\eta} k^\lambda \p_\lambda \, {\cal A}^\nu_{\rm eq} (x,k) = 0 \qquad \Rightarrow \p_\lambda S^{\lambda , \mu \nu }_{\rm GLW}(x) = 0
\eel
\section{Formulation of perfect-fluid hydrodynamics with spin using classical treatment of spin degrees of freedom}
In this section we discuss the formulation of  perfect-fluid hydrodynamics with spin using the classical treatment of spin.
\label{sec:3}
\subsection{Classical spin dependent equilibrium distribution function}
\label{subsec:3.1}
In the classical treatments of particles with spin-1/2 one introduces internal angular momentum  tensor of particles $s^{\alpha\beta}$~\cite{Mathisson:1937zz} which is connected by  two orthogonal four vectors namely the particle four-momentum $p_\gamma$ and spin four-vector~$s_\delta$~\cite{Itzykson:1980rh} by following relation,
\bea
s^{\alpha\beta} = \f{1}{m} \epsUabgd p_\gamma s_\delta.
\label{eq:salbe}
\eea
%
Further, from Eq. (\ref{eq:salbe}) we can obtain,
\bea
s^{\alpha} = \f{1}{2m} \epsUabgd p_\beta s_{\gamma \delta}.
\label{eq:sabinv}
\eea
Projecting four-vector $s^{\alpha}$ by four momentum $p^{\alpha}$ we can get  $p \cdot s = 0$ {\it i.e.} spin four vector and particle four momentum satisfy the orthogonality relation.
In particle rest frame (PRF), particle four momentum is given by $p^\mu = (m,0,0,0)$. The condition $p_\alpha s^{\alpha}  = 0$ implies that the spin four-vector $s^\alpha$ has only spatial components {\it i.e.}, $s^\alpha = (0,\sv_*)$ with the normalization $|\sv_*|=\mathfrak{s}$. The length
of the spin vector given by  $-s^2={|\sv_*|^2} ={\mathfrak{s^2}}= \f{1}{2} \left( 1+ \f{1}{2}  \right)=\f{3}{4}$.

By identifying the so-called collisional invariants of the Boltzmann equation, following the equilibrium distribution function $f^\pm_{s, \rm eq}(x,\pv,s)$ for particles and antiparticles with spin-1/2 can be constructed~ \cite{Florkowski:2018fap,Bhadury:2020puc},
\begin{equation}
f^\pm_{s, \rm eq}(x,\pv,s) =  f_{\rm eq}^{\pm}(x,\pv)\exp\left( \frac{1}{2} \omega_{\mu\nu} s^{\mu\nu} \right).
\label{eq:feqxps}
\end{equation}
In the above equation $f_{\rm eq}^{\pm}(x,\pv)=\exp\left[-p^\mu\beta_\mu(x)\pm\xi(x)\right]$ is the J\"uttner distribution function. The tensor $\omega_{\alpha \beta}(x)$ is the polarization tensor as introduced in subsection \ref{subsec:2.1}. In this formalism, it plays a role similarly to the chemical potential conjugate to the spin angular momentum. 

 It is important to note that in this approach $S^{\mu\nu}$ and $\omega^{\mu\nu}$ are dimensionless which are measured in units of $\hbar$.
Ordinary equilibrium phase space-distribution function can be obtained by the normalizing $f^\pm_{s, \rm eq}(x,\pv,s)$ in the following way
\begin{equation}
\int dS \, f^\pm_{s, \rm eq}(x,\pv,s) = f^\pm_{\rm eq}(x,\pv), 
\label{a}
\end{equation}
where $dS = (m/\pi {\mathfrak{s}}) \,  d^4s \, \delta(s \cdot s + {\mathfrak{s}}^2) \, \delta(p \cdot s)$.
\subsection{Procedure to formulate perfect-fluid hydrodynamics with spin}
\label{subsec:3.2}
We need to calculate the conserved charge current, energy momentum and spin tensors. The structures of hydrodynamic quantities, $N^\mu$, $T^{\mu\nu}$ and $S^{\lambda, \mu\nu}$ in the standard kinetic theory description are well known and are connected to the behaviour of the microscopic constituents of the system via a phase-space distribution function  $f_{\rm eq}(x,p,s)$.

Using the above equilibrium distribution function $f_{\rm eq}(x,p,s)$ (\ref{eq:feqxps}) the hydrodynamic quantities such as charge current, Energy-momentum tensor and the Spin tensor can be obtained as follows.
\subsection{Charge current}
\label{subsubsec:3.3}
The charge current can be obtained from the standard definition
\bea
N^\mu_{\rm eq} &=& \int dP   \int dS \, \, p^\mu \, \left[f^+_{\rm eq}(x,p,s)-f^-_{\rm eq}(x,p,s) \right],
\label{eq:Neq-sp0}
\eea
Using the equilibrium functions (\ref{eq:feqxps}) we get
\bea
N^\mu_{\rm eq} = 2 \sinh(\xi) \int dP \, p^\mu \, \exp\LB- p \cdot \beta \RB
\int dS \, \exp\LB \f{1}{2}  \omega_{\alpha \beta} s^{\alpha\beta} \RB.
\label{eq:Neq-sp1}
\eea
Note that due to inconsistency of the classical description~\cite{Florkowski:2018fap} with semi-classical Wigner-function~\cite{Florkowski:2018ahw,Florkowski:2018fap,Florkowski:2019qdp} at arbitrary large values of the polarization tensor~\cite{Florkowski:2018fap} we consider the case of small values of the polarization tensor $\omega$, in this case the  exponential function with $\omega^{\mu\nu}$ can be expanded upto linear order in $\omega$,
\beq
N^\mu_{\rm eq} &=& 2 \sinh(\xi) \int dP \, p^\mu \, e^{- p \cdot \beta} \int dS \, \left[1 +  \f{1}{2}  \omega_{\alpha \beta} s^{\alpha\beta}\right] 
\label{eq:Neq-sp21}
\eeq
After carrying out integration first over spin and then momentum we get
\beq
N^\alpha_{\rm eq} = n u^\alpha, \label{eq:ccc}
\eeq
where 
\beq
n = 4 \, \sinh(\xi)\, \n0(T) \label{eq:nd}
\eeq
is the net charge density~\cite{Florkowski:2017ruc} with $\n0(T)$ defined above in Eq.~(\ref{eq:N0}). We note here that the above expression for charge current (Eq. (\ref{eq:ccc})) agrees with Eq. (\ref{eq:cc}) in the small spin polarization limit {\it i.e.} $\zeta<1$.
\subsection{Energy-momentum tensor}
\label{subsubsec:3.4}
The energy-momentum tensor can be obtained by taking the second moment of distribution function (\ref{eq:feqxps}) in momentum space
\bea
T^{\mu \nu}_{\rm eq}&=& \int dP   \int dS \, \, p^\mu p^\nu \, \left[f^+_{\rm eq}(x,p,s) + f^-_{\rm eq}(x,p,s) \right] 
\label{eq:Teq-sp02}
\eea
Substituting $f^\pm_{s, \rm eq}(x,\pv,s)$, from (\ref{eq:feqxps}) in above equation we get
\bea
T^{\mu \nu}_{\rm eq}
&=& 2 \cosh(\xi) \int dP \, p^\mu p^\nu \, \exp\LB - p \cdot \beta\RB
\int dS \, \exp\LB  \f{1}{2}  \omega_{\alpha \beta} s^{\alpha\beta} \RB,
\label{eq:Teq-sp03}
\eea
Now if we consider small $\omega$ limit and carry out integration over spin and momentum we can get, 
\bel{Tmn}
T^{\alpha\beta}_{\rm eq}(x) &=& \varepsilon u^\alpha u^\beta - P \Delta^{\alpha\beta}. \label{eq:tmunucl}
\eel
where
\bel{enden}
\varepsilon = 4 \,\cosh(\xi) \, \e0(T)
\eel
and
\bel{prs}
P = 4 \, \cosh(\xi) \, \P0(T),
\eel
are the net energy density and pressure respectively~\cite{Florkowski:2017ruc}.  The auxiliary objects, $\e0(T)$, $\P0(T)$ are same as given in Eqs. (\ref{enden0}) and (\ref{prs0}). 
Similar to case of charge current, it can be easily noticed here that the expression (\ref{eq:tmunucl}) for $T^{\mu \nu}_{\rm eq}$ obtained here agrees with $\TmnU_{\rm GLW}$ in small polarization limit. 
\subsection{Spin tensor}
\label{subsubsec:3.5}
The spin tensor is defined as follows,
\bea
S^{\lambda, \mu\nu}_{\rm eq} &=& \int dP   \int dS \, \, p^\lambda \, s^{\mu \nu} 
\left[f^+_{\rm eq}(x,p,s) + f^-_{\rm eq}(x,p,s) \right] \nn \\
&=& 2 \cosh(\xi) \int dP \, p^\lambda \, \exp\LB - p \cdot \beta \RB
\int dS \, s^{\mu \nu} \, \exp\LB  \f{1}{2}  \omega_{\alpha \beta} s^{\alpha\beta} \RB.
\label{eq:Seq-sp01}
\eea
In the leading-order approximation in $\omega$ integration on spin variable can be performed and we get 
\bea
S^{\lambda, \mu\nu}_{\rm eq} 
&=& \f{4}{3 m^2} {{\mathfrak{s}}}^2 \cosh(\xi) \int dP \, p^\lambda \, e^{ - p \cdot \beta }
\LB m^2 \omnU + 2 p^\alpha p^{[\mu} \omega^{\nu ]}_{\,\,\alpha} \RB. 
\label{eq:Seq-sp1}
\eea
Now carrying out the momentum integration we obtain
\bea
S^{\lambda, \mu\nu}_{\rm eq}
&=&  {\cal C} \left( n_{(0)}(T) u^\lambda \omega^{\mu\nu}  +  S^{\lambda , \mu \nu }_{\Delta\GLW} \right).
\label{eq:Smunulambda_de_Groot3}
\eea 
In the above expression $
{\cal C}= \f{4}{3} {{\mathfrak{s}}}^2 \cosh{\xi}$, while $S^{\lambda , \mu \nu }_{\Delta\GLW}$ is given by Eq.(\ref{SDeltaGLW}). Note that for spin-1/2 particles ${{\mathfrak{s}}}^2=3/4$, therefore,  $S^{\lambda, \mu\nu}_{\rm eq}$ is matches with the one obtained using the Wigner function approach in the small polarization limit (see Eq.(\ref{eq:Smunulambda_de_Groot2}). 
%
%
\subsection{Entropy Current}
\label{subsubsec:3.6}
To obtain the conserved entropy current we adopt the Boltzmann definition as follows
\bea
H^\mu = - \int dP \int dS \, p^\mu 
\LSB 
f^+_{\rm eq} \LB \ln f^+_{\rm eq} -1 \RB  + 
f^-_{\rm eq} \LB \ln f^-_{\rm eq} -1 \RB \RSB.
\label{eq:H1}
\eea 
Using Eq.(\ref{eq:feqxps}),  (\ref{eq:Neq-sp0}), (\ref{eq:Teq-sp02}) and  (\ref{eq:Seq-sp01}) following expression for entropy current can be obtained
\bea
H^\mu = \beta_\alpha T^{\mu \alpha}_{\rm eq} -\f{1}{2} \omega_{\alpha\beta} S^{\mu, \alpha \beta}_{\rm eq}
-\xi N^\mu_{\rm eq} + {\cal N}^\mu_{\rm eq}
\label{eq:H2}
\eea 
where
\bea
{\cal N}^\mu_{\rm eq} = \f{\cosh(\xi)}{\sinh(\xi)}  N^\mu_{\rm eq}.
\label{eq:NcalN}
\eea 
Taking the partial derivative of above Eq.(\ref{eq:H2}) and using the conservation laws for charge current, energy-momentum tensor and spin tensor we obtain 
\bea
\p_\mu H^\mu = \left( \p_\mu \beta_\alpha \right) T^{\mu \alpha}_{\rm eq} 
-\f{1}{2} \left( \p_\mu \omega_{\alpha\beta} \right) S^{\mu, \alpha \beta}_{\rm eq}
- \left(\p_\mu \xi \right)  N^\mu_{\rm eq} + \p_\mu {\cal N}^\mu_{\rm eq}.
\label{eq:H3}
\eea 
Now using Eqs. (\ref{eq:Neq-sp0}), (\ref{eq:Teq-sp03}), (\ref{eq:Seq-sp01}), (\ref{eq:NcalN}) and applying the conservation of charge current it can be easily shown that right hand side of Eq.~\eqref{eq:H3} vanishes {\it i.e.} four-entropy current is conserved,
\bea 
\p_\mu H^\mu = 0.
\label{eq:entcon}
\eea
%

\section{Dissipative effects in relativistic hydrodynamics with spin}
\label{sec:4}
In the previous sections we presented well established version of the perfect-fluid hydrodynamics with spin in two different ways namely: Wigner function approach and a approach based on the classical treatment of spin degrees of freedom. In this section we present the results of our article~\cite{Bhadury:2020cop} where we include dissipative effects by using classical relaxation time approximation (RTA) for the collision terms in the classical kinetic equations as discussed in Ref~\cite{Bhadury:2020puc}.

\subsection{Classical kinetic equation for particles with spin-1/2 in RTA}
\label{sec:4.1}
We consider the case when mean fields are absent, in this case the distribution functions (\ref{eq:feqxps}) obey the following classical equation 
\begin{equation}\label{RTA_spin}
p^\mu \partial_\mu f^\pm_s(x,\pv,s) =C[f^\pm_s(x,\pv,s)]
\end{equation}
where $C[f^\pm_s(x,\pv,s)]$ is accounted for the effect of collisions. In RTA $C[f^\pm_s(x,\pv,s)]$ is given by
\begin{equation}
C[f^\pm_s(x,\pv,s)] =  p \cdot u
\, \frac{f^\pm_{s, \rm eq}(x,\pv,s)-f^\pm(x,\pv,s)}{\tau_{\rm eq}},
\label{eq:col}
\end{equation}
Now we expand single particle distribution around its equilibrium value in powers of space-time gradients,
\begin{equation}
f^\pm_s(x,\pv,s) =f^\pm_{s, \rm eq}(x,\pv,s)+\delta f^\pm_s(x,\pv,s),
\label{cha-ensk}
\end{equation}
Substituting Eq.(\ref{cha-ensk}) and (\ref{eq:col}) in Eq.(\ref{RTA_spin}) we find
\begin{equation}\label{RTA_spin2}
p^\mu \partial_\mu f^\pm_{s, \rm eq}(x,\pv,s) =-p \cdot u
\, \frac{\delta f^\pm_s(x,\pv,s)}{\tau_{\rm eq}},
\end{equation}
Using the expressions for the equilibrium distribution functions (\ref{eq:feqxps}) in linear order of $\omega$ in Eq.(\ref{RTA_spin2}) we get
\begin{equation}\label{del feq lim}
\delta f^\pm_s =- \frac{\tau_{eq}}{(u \cdot p)}e^{\pm\xi - p\cdot \beta} \bigg[ \Big(\pm p^\mu\partial_\mu\xi - p^\lambda p^\mu\partial_\mu\beta_\lambda\Big) \bigg(1 + \frac{1}{2}s^{\alpha\beta}\omega_{\alpha\beta}\bigg) + \frac{1}{2} p^\mu s^{\alpha\beta}(\partial_\mu\omega_{\alpha\beta})\bigg]
\end{equation}
Dissipative effects in the conserved quantities are induced by $\delta f^\pm_s$. 
Before we proceed to calculate the dissipative corrections to charge-current, energy-momentum tensor and spin tensor, we discuss how kinetic description retains conservation law. To see this we take the similar moments of kinetic equations (\ref{RTA_spin})  as defined by Eqs. (\ref{eq:Neq-sp0}), (\ref{eq:Teq-sp02}) and (\ref{eq:Seq-sp01}) which gives,
\beq
\p_\mu N^{\mu}(x)&=&-u_{\mu}\left(\frac{N^{\mu}(x)-N^{\mu}_{\rm eq}(x)}{\tau_{\rm eq}}\right), \label{eq:cc2}\\
\p_\mu T^{\mu\nu}(x)&=&-u_{\mu}\left(\frac{T^{\mu\nu}(x)-T^{\mu\nu}_{\rm eq}(x)}{\tau_{\rm eq}}\right), \label{eq:emt2}\\
\p_\lambda S^{\lambda , \mu \nu}(x)&=&-u_{\lambda}\left(\frac{S^{\lambda , \mu \nu}(x)-S^{\lambda , \mu \nu}_{\rm eq}(x)}{\tau_{\rm eq}}\right).\label{eq:st2}
\eeq
From the above equations, in order to have conserve charge current ($\p_\mu N^{\mu}=0$), energy-momentum tensor ($\p_\mu T^{\mu\nu}=0$), and spin tensor ($\p_\lambda S^{\lambda , \mu \nu}=0$) we must have
\beq
u_{\mu}\delta N^{\mu}&=&0,\label{eq:lm1}\\
u_{\mu}\delta T^{\mu\nu}&=&0,\label{eq:lm2}\\
u_{\lambda}\delta S^{\lambda , \mu \nu}&=& 0, \label{eq:lm3}
\eeq
Eqs.~(\ref{eq:lm1}), (\ref{eq:lm2}), and (\ref{eq:lm3}) are known as the Landau matching conditions. In these equations, $\delta N^\mu$, $\delta T^{\mu\nu}$, and $\delta S^{\lambda,\mu\nu}$ are the dissipative corrections to charge current, energy momentum  and spin tensor which are defined in terms non-equilibrium corrections of the distribution functions as follows
\begin{align}
&\delta N^\mu = \int \mathrm{dP}~\mathrm{dS}~p^\mu (\delta f^+_s-\delta f^-_s)\label{eq:dN}\\
&\delta T^{\mu\nu} = \int \mathrm{dP}~\mathrm{dS}~p^\mu p^\nu (\delta f^+_s+\delta f^-_s)\label{eq:dT}\\
&\delta S^{\lambda,\mu\nu}=\int \mathrm{dP~dS}~p^{\lambda} s^{\mu\nu} (\delta f^+_s+\delta f^-_s).\label{eq_dspin}
\end{align}
The conserved quantities that are obtained from the moments of the transport equations (\ref{RTA_spin}) can be decomposed in terms of the hydrodynamic degrees of freedom. The decomposition of charge current can be done as follows
\beq
N^\mu &=&\int {dP~dS}~p^{\mu} \left[f^+(x, p, s) - f^-(x, p, s)\right]\nn\\&=&N^\mu_{\rm eq}+\delta N^{\mu}=nu^{\mu}+\nu^\mu\label{Nmu1}.
\eeq
Here, the quantity $\nu^\mu$ is known as the particle diffusion current.
The energy-momentum tensor can be decomposed as 
\bea
T^{\mu \nu}&=&\int {dP~dS}~p^{\mu} p^{\nu} \left[f^+(x, p, s) + f^-(x, p, s)\right]\nn\\
&=& T^{\mu \nu}_{\rm eq}+\delta T^{\mu \nu} \nn \\
&=& \varepsilon u^\mu u^\nu - P \Delta^{\mu\nu}+\pi^{\mu\nu}-\Pi \Delta^{\mu\nu}. \label{Tmunu1}
\eea
In this decomposition, the dissipative quantities $\pi^{\mu\nu}$, and $\Pi$ are known as shear stress tensor, and bulk pressure. Here, we not that this decomposition is done in the Landau frame, where $T^{\mu\nu}u_{\nu}=\varepsilon u^\mu$. 
Disipative corrections to the spin tensor are given by following decomposition
\beq
S^{\lambda,\mu\nu} &=&\int \mathrm{dP~dS}~p^{\lambda} s^{\mu\nu}\left[f^+(x, p, s) + f^-(x, p, s)\right]\nn\\&=&S^{\lambda,\mu\nu}_{\rm eq}+\delta S^{\lambda,\mu\nu}.\label{Slmunu}
\eeq
The non-equilibrium charge density $n$, energy density $\varepsilon$, and pressure $P$ can be obtained by the Landau matching conditions which implies that {\it i.e.} at local thermodynamic equilibrium we must have, $n =n_{\rm {eq}} = u_{\mu } N_{\rm {eq}}^{\mu }$, and $\varepsilon = \varepsilon_{\rm {eq}} = u_{\mu }u_{\nu} T^{\mu\nu}_{\rm {eq}}$ and $P = P_{\rm {eq}} = -\frac{1}{3}\Delta_{\mu\nu} T^{\mu\nu}_{\rm {eq}}$. It is important to note that the choice of Landau frame and matching conditions enforces the following constraints on the dissipative currents 
\beq
u_\mu \nu^\mu&=&0, \nn\\
u_\mu \pi^{\mu\nu}&=&0. \label{matching_conditions}
\eeq		 
\subsection{Evaluation of dissipative quatities}
\label{4.6}
 Using the Eqs.~(\ref{Tmunu1}) and (\ref{Nmu1}) and then writing conservation laws of energy-momentum ($\partial_\mu T^{\mu \nu}= 0$) and particle four flow ($\partial_\mu N^{\mu \nu}= 0$), following equations that dictates space-time  evolution of various thermodynamic quantities such as tempereture $T$, chemical potential $\mu$ and flow variable $u^{\mu}$ can be obtained
\begin{align}
&\dot{\epsilon} + (\epsilon + P + \Pi)~\theta -\pi^{\mu\nu}\sigma_{\mu\nu} = 0\label{eq:con_e}\\
&(\epsilon + P) \dot{u}^\alpha - \nabla^\alpha P + \Delta^\alpha_\mu \partial_\mu \pi^{\mu\nu} = 0\label{eq:con_P}\\
&\dot{n} + n\theta + \partial_\mu \nu^\mu = 0\label{eq:con_n}
\end{align}
In the above equations we have used the notations, $\dot{\varepsilon}=D{\varepsilon}$,  $\dot{n}=D{n}$ and $\dot{u}^\alpha=D{u}^\alpha$ with  $D=u^{\mu}\partial_{\mu}$ being the convective derivative. In addition we have defined $\theta=\partial_{\mu}u^{\mu}$ which is known as the expansion scalar and $\nabla^\mu = \Delta^{\mu\nu} \partial_\nu$ the transverse gradient. Notation, $\sigma ^{\mu\nu}=\frac{1}{2}\left(\nabla ^{\mu }u^{\nu }+\nabla ^{\nu }u^{\mu}\right)-\frac{1}{3}\Delta ^{\mu\nu}\left.(\nabla ^{\lambda }u_{\lambda }\right)$ is used for the shear flow tensor. 
In addition, we also have conservation of spin tensor 
\begin{align}
\partial_\lambda S^{\lambda,\mu\nu}  = 0 \label{eq:con_S}
\end{align}
Keeping the terms upto first order in velocity gradients in above Eqs.(\ref{eq:con_e}), (\ref{eq:con_P}), (\ref{eq:con_n}) and (\ref{eq:con_S}) and using Eqs.(\ref{eq:nd}), (\ref{enden}), (\ref{prs}) and (\ref{eq:Smunulambda_de_Groot3}) we get

\beq
\dot{\xi} &=&\xi_\theta \, \theta\label{eq:xidot1} \\
\dot{\beta}&=& \beta_\theta \, \theta \label{eq:betadot1}\\
\beta  \dot{u}^{\alpha }&=&\frac{n_0 \tanh (\xi )}{\left(\varepsilon _0+P_0\right)}\left(\nabla ^{\alpha }\xi \right)-\left(\nabla ^{\alpha }\beta \right). \label{eq:udot1}\\
\dot{\omega }^{\mu \nu}&=& D_{\Pi }^{\mu \nu }\theta +\left(\nabla ^{\alpha }\xi \right)D_n^{[\mu \nu ]}{}_{\alpha }+D_{\pi }^{[\nu }{}_{\lambda }\sigma ^{\lambda\mu ]}+D_{\text{$\Sigma $1}}^{\alpha }\nabla ^{[\mu }\omega ^{\nu ]}{}_{\alpha }+D_{\text{$\Sigma $2}}^{[\mu \nu ]\alpha }\nabla ^{\lambda }\omega _{\alpha \lambda } \label{eq:dotomega}
\eeq
where
\beq
\xi_\theta &=&\frac{\sinh (\xi ) \cosh (\xi ) \left(\varepsilon _0^2-n_0 T \left(\left(3+z^2\right)P_0+2 \varepsilon _0\right)\right)}{n_0 T \cosh ^2(\xi ) \left( \left(3+z^2\right)P_0+3 \varepsilon _0\right)-\varepsilon _0^2 \sinh ^2(\xi )}\nn\\ \label{eq:xith} \\
\beta_\theta&=& \frac{n_0 \left(\cosh ^2(\xi )P_0 +\varepsilon _0\right)}{n_0 T \cosh ^2(\xi ) \left(\left(3+z^2\right)P_0+3 \epsilon _0\right)-\varepsilon _0^2 \sinh ^2(\xi )}. \label{eq:Dmunu}
\eeq
The explicit expressions for various $D$-coefficients appearing in Eq. (\ref{eq:dotomega}) are provided in Appendix \ref{AD3}.
Here we would like to point out out that while deriving the dynamical equation (\ref{eq:dotomega}) we have eliminated a term $u_{\nu}\dot{\omega }^{\mu \nu }$ from the inital expression of $\dot{\omega }^{\mu \nu }$. The term $u_{\nu}\dot{\omega }^{\mu \nu }$ is given by a dynamical equation which is obtained by taking projection of the initial expression of $\dot{\omega}^{\mu\nu}$ along $u_\nu$.
\beq
u_{\nu }\dot{\omega }^{\mu \nu }&=&C_{\Pi }^{\mu }\theta +C_{n{}{\lambda }}^{\mu }{}(\nabla ^{\lambda }\xi )+C_{\pi{}{\alpha }}\sigma ^{\alpha \mu }+C_{\text{$\Sigma$} {}{\nu }}^{\mu } \nabla _{\lambda }\omega ^{\nu \lambda }\nn
\eeq
The various $C$-coefficients in above equation are given in Appendix \ref{AC3}.

The dissipative forces are the result of non-zero gradients in the system. In the  present case we have restricted ourself only to first order in gradients. In this case, the shear stress $(\pi^{\mu\nu})$, bulk viscous pressure ($\Pi$) and particle diffusion current $(n^\mu)$ can be obtained using $\delta T^{\mu\nu}$ and $\delta N^\mu$ as follows
\begin{align}
&\pi^{\mu\nu}= \Delta^{\mu\nu}_{\alpha\beta}\delta T^{\alpha\beta} = \Delta^{\mu\nu}_{\alpha\beta} \int \mathrm{dP}~\mathrm{dS}~p^\alpha p^\beta (\delta f^+_s+\delta f^-_s)\label{eq:shear}\\
&\Pi= -\frac{1}{3}\Delta_{\alpha\beta}\delta T^{\alpha\beta} = - \frac{1}{3}\Delta_{\alpha\beta} \int \mathrm{dP}~\mathrm{dS}~p^\alpha p^\beta (\delta f^+_s+\delta f^-_s)\label{eq:bulk}\\
&\nu^\mu = \Delta^\mu_\alpha~\delta N^\alpha = \Delta^\mu_\alpha \int \mathrm{dP}~\mathrm{dS}~p^\alpha (\delta f^+_s-\delta f^-_s)\label{eq:nu}
\end{align}
In a similar way, dissipative part of the spin-tensor is given by
\begin{align}
\delta S^{\lambda,\mu\nu} &=\int \mathrm{dP~dS}~p^{\lambda} s^{\mu\nu} (\delta f^+_s+\delta f^-_s).\label{eq_dspin1}
\end{align}
Using Eq.~(\ref{del feq lim}) in Eqs.(\ref{eq:shear}), (\ref{eq:bulk}) and (\ref{eq:nu}) carrying out the integration on spin and momentum the dissipative quatities up to first order order in gradient are found be
\begin{align}
\pi^{\mu\nu} = 2 \tau_{eq}\,\beta_\pi \sigma^{\mu\nu}, \quad \Pi = - \tau_{eq}\,\beta_\Pi \theta, \quad n^\mu = \tau_{eq}~\beta_n \nabla^{\mu} \xi
\end{align}
where $\beta_\pi$, $\beta_\Pi$ and $\beta_n$ are the first order transport coefficients for massive particle with finite chemical potential. These coefficients are given below
\begin{align}
&\beta_\pi = 4~I^{(1)}_{42} \cosh({\xi})\\
&\beta _{\Pi }=4\Bigg[\frac{n_0 \left(\cosh (\xi ) {\sinh}^2(\xi ) \left(\varepsilon _0 \left(P_0+\varepsilon _0\right)-n_0 T \left(P_0 \left(z^2+3\right)+3 \varepsilon _0\right)\right)\right)}{\beta  \left(\varepsilon _0^2 \sinh ^2(\xi )-n_0 T \cosh ^2(\xi ) \left(P_0 \left(z^2+3\right)+3 \varepsilon _0\right)\right)}\nn\\
&~~~-\frac{\cosh (\xi )}{\beta } \left(\frac{n_0 \left(P_0+\varepsilon _0\right) \left(P_0 \cosh ^2(\xi )+\varepsilon _0\right)}{n_0 T \cosh ^2(\xi ) \left(P_0 \left(z^2+3\right)+3 \varepsilon _0\right)-\varepsilon _0^2 \sinh ^2(\xi )}\right)+\frac{5\beta }{3}I^{(1)}_{42}\Bigg]\\
&\beta_n = 4 \bigg[\bigg(\frac{n_0 \tanh(\xi)}{\varepsilon_0 + P_0}\bigg)~I^{(0)}_{21}~(\sinh{\xi}) - I^{(1)}_{21}~(\cosh{\xi})\bigg]. 
\end{align}
Dissipative corrections in the spin tensor can be found by substituting  Eq.(\ref{del feq lim}) in (\ref{eq_dspin1}) and then carrying out integration over spin and momentum variables, 	
\begin{eqnarray}
\delta S^{\lambda,\mu\nu} &=& \tau_{\rm eq} \Big[ \beta^{\lambda,\mu\nu}_{\Pi}\, \theta + \beta^{\kappa\lambda,\mu\nu}_{n}\, (\nabla_\kappa \xi) + \beta^{\lambda\kappa\delta,\mu\nu}_{\pi}\, \sigma_{\kappa\delta}
+ \beta^{\kappa\lambda\alpha\beta,\mu\nu}_{\Sigma}\, (\nabla_\kappa \omega_{\beta\alpha}) \Big]. 
\label{deltaS}
\end{eqnarray}
In the above Eq.~(\ref{deltaS}) different $\beta$-coefficients appearing on the are the knetic coefficients for spin. Explicit form of these coefficients are given below. 
\begin{eqnarray}
\beta^{\lambda,\mu\nu}_{\Pi} &=&
\beta_{\Pi }^{(1)}u^{[\mu }\omega ^{\nu ]\lambda } + \beta_{\Pi }^{(2)}u^{\lambda }u^{\alpha }u^{[\mu }\omega ^{\nu ]}{}_{\alpha } +
\beta_{\Pi }^{(3)}\Delta ^{\lambda [\mu }u_{\alpha }\omega ^{\nu ]\alpha }
\label{eq:betaPi}\\
\beta_{\pi}^{\lambda \kappa \delta ,\mu \nu }&=&\beta_{\pi }^{(1)}\Delta^{[\mu \kappa }\Delta^{\lambda \delta }u_{\alpha }\omega ^{\nu ]\alpha }+\beta_{\pi }^{(2)}\Delta^{\lambda \delta }u^{[\mu }\omega ^{\nu ]\kappa }+\beta_{\pi }^{(3)}u^{[\mu }\Delta^{\nu ]\delta }\Delta^{\lambda}_{\alpha}\omega ^{\alpha \kappa }\nn\\&+&\beta_{\pi }^{(4)}\Delta ^{\lambda [\mu }\omega ^{\rho \kappa } u_{\rho }\Delta^{\nu ]\delta }
\label{eq:betapi}\\
\beta_n^{\lambda \kappa ,}{}^{\mu \nu }&=&\beta_n^{(1)}\Delta ^{\lambda \kappa } \omega ^{\mu \nu }+\beta_n^{(2)}\Delta ^{\lambda \kappa }u^{\alpha }u^{[\mu }\omega ^{\nu ]}{}_{\alpha }+\beta_n^{(3)}\Delta ^{\lambda \alpha }\Delta ^{[\mu \kappa }\omega ^{\nu ]}{}_{\alpha }\nn\\&+&\beta_n^{(4)}u^{[\mu }\Delta ^{\nu ]\kappa }u^{\rho }\omega ^{\lambda }{}_{\rho }+\beta_n^{(5)}\Delta ^{\lambda [\mu }\omega ^{\nu ]\kappa }+\beta_n^{(6)}\Delta ^{\lambda [\mu }u^{\nu ]}u_{\alpha }\omega ^{\alpha \kappa}
\label{eq:betan}\\
\beta_{\Sigma }^{\eta \beta \gamma \lambda ,\mu \nu }&=&\beta_{\Sigma}^{(1)}\Delta^{\lambda \eta }g^{[\mu \beta }g^{\nu ]\gamma }+\beta_{\Sigma }^{(2)} u^{\gamma }\Delta ^{\lambda \eta }u^{[\mu }\Delta ^{\nu ]\beta }\nn\\&+&\beta_{\Sigma}^{(3)}\left(\Delta ^{\lambda \eta }\Delta ^{\gamma [\mu }g^{\nu ]\beta }+\Delta ^{\lambda \gamma }\Delta ^{[\mu \eta }g^{\nu ]\beta }+\Delta ^{\gamma \eta }\Delta ^{\lambda [\mu }g^{\nu ]\beta }\right)\nn\\&+&\beta_{\Sigma}^{(4)}\Delta ^{\gamma \eta }\Delta ^{\lambda [\mu }\Delta^{\nu ]\beta}+\beta_{\Sigma}^{(5)}u^{\gamma }\Delta ^{\lambda \beta }u^{[\mu }\Delta ^{\nu ]\eta }\label{eq:betaSig}
\end{eqnarray}
where, the scalar  $\beta$--coefficients are given in Appendix (\ref{AB3}).
\section{Application to heavy ion collisions}
\label{sec:5}
In this section we report recently discussed \cite{Florkowski:2018fap,Florkowski:2019qdp} application of perfect-fluid hydrodynamics with spin to determine physical observable which describe the spin polarization of particles.
\subsection{Boost-invaraint evolution equations of perfect-fluid hydrodynamics with spin}
\label{subsec:5.1}
We consider the transversely homogeneous and boost-invariant longitudinal expansion of the collision firewall, also known as Bjorken flow  \cite{Bjorken:1982qr}. In this case, it is easy to use following four-vector basis
\beq
u^\alpha &=& \frac{1}{\tau}\LR t,0,0,z \RR = \LR \cosh\eta, 0,0, \sinh\eta\RR, \nn \\
X^\alpha &=& \LR 0, 1,0, 0 \RR,\nn\\
Y^\alpha &=& \LR 0, 0,1, 0 \RR, \nn\\
Z^\alpha &=& \frac{1}{\tau}\LR z,0,0,t \RR = \LR \sinh\eta, 0,0, \cosh\eta \RR.
\lab{BIbasis}
\eeq
Here, variables $\tau = \sqrt{t^2-z^2}$ and $\eta =  \ln((t+z)/(t-z))/2$  are the longitudinal proper time and space time rapidity. 
The basis vector $u^\alpha$ is a time like vector which is normalized to unity. The other basis vectors, $X^\alpha$, $Y^\alpha$ and $Z^\alpha$ are space-like and orthogonal to $u^\alpha$ as well as to each other.

Spin polarization tensor $\omega^{\mu\nu}$ is antisymmetric and can be decomposed in terms of two new four-vectors $\kappa^{\mu}$ (electric-like) and $\omega^{\mu}$ (magnetic-like) with respect to flow four-vector $u^{\mu}$  
\beq
\omega^{\mu\nu}&=&\kappa^{\mu}u^{\nu}-\kappa^{\nu}u^{\mu}+\epsilon_{\mu\nu\alpha\beta}u^{\alpha}\omega^{\beta}. \label{eq:sd}
\eeq
The four vectors  $\kappa^{\mu}$  and $\omega^{\mu}$ satisfy the following orthogonality conditions with $u^{\mu}$.
\beq
\kappa\cdot{u}=0, \quad \omega\cdot{u}=0 \label{eq:oc}
\eeq
Using the basis (\ref{BIbasis}) and the orthogonality conditions (\ref{eq:oc}), four-vectors $\kappa^{\mu}$  and $\omega^{\mu}$ can be decomposed as follows
\beq
\kappa^{\mu}&=&C_{\kappa X} X^{\mu}+C_{\kappa Y} Y^\mu+C_{\kappa Z} Z^\mu \label{eq:kd} \\
\omega^{\mu}&=&C_{\omega X} X^{\mu}+C_{\omega Y} Y^\mu+C_{\omega Z} Z^\mu \label{eq:od}
\eeq
where due to boost invariant notion all the scalar coefficients (${C}'s$) are functions of proper time only. 

Substituting Eqs. (\ref{eq:kd}) and (\ref{eq:od}) in Eq. (\ref{eq:sd}) we can represent the spin polarization tensor in terms of boost-invariant basis as follows~\cite{Florkowski:2019qdp}), 
\beq
\omega_{\mu\nu} &=& C_{\kappa Z} (Z_\mu u_\nu - Z_\nu u_\mu) 
 + C_{\kappa X} (X_\mu u_\nu - X_\nu u_\mu)  \nonumber \\
&& + C_{\kappa Y} (Y_\mu u_\nu - Y_\nu u_\mu)\label{eq:omegamunu} \\
&& + \, \epsilon_{\mu\nu\alpha\beta} u^\alpha (C_{\omega Z} Z^\beta + C_{\omega X} X^\beta + C_{\omega Y} Y^\beta). \nn
\eeq
Using the boost boost-invariant decomposition of $\omega_{\mu\nu}$ in  (\ref{eq:kineqSC1}) and taking projections by $u_{\mu}X_{\nu}$,$u_{\mu}Y_{\nu}$,$u_{\mu}Z_{\nu}$,$Y_{\mu}Z_{\nu}$,$X_{\mu}Z_{\nu}$,$X_{\mu}Y_{\nu}$ the form of conservation laws for spin tensor can be written in terms of the following six evolution equations of coefficients ${C}'s$
\beq
{\rm diag}(\cal{L},\cal{L},\cal{L},\cal{P},\cal{P},\cal{P}){~\bm{\dot{C}}}&=&{\rm diag}({\cal{Q}}_1,{\cal{Q}}_1,{\cal{Q}}_2,{\cal{R}}_1,{\cal{R}}_1,{\cal{R}}_2){~\bm{C}} \label{eq:C-coeff}
\eeq
where $\bm{C}=(C_{\kappa X},C_{\kappa Y},C_{\kappa Z},C_{\omega X},C_{\omega Y},C_{\omega Z})$ and  ${~\bm{\dot{C}}}=u^{\mu}\partial_{\mu}=\partial_{\tau}$ while
\beq
{\cal L}(\tau)&=&{\cal A}_1-\frac{1}{2}{\cal A}_2-{\cal A}_3,\nn\\
{\cal P}(\tau)&=&{\cal A}_1,\nn \\
{\cal{Q}}_1(\tau)&=&-\left[\dot{{\cal L}}+\frac{1}{\tau}\left({\cal L}+ \frac{1}{2}{\cal A}_3\right)\right],\nn\\
{\cal{Q}}_2(\tau)&=&-\left(\dot{{\cal L}}+\frac{{\cal L}}{\tau} \right),\nn\\
{\cal{R}}_1(\tau)&=&-\left[\Dot{\cal P}+\frac{1}{\tau}\left({\cal P} -\frac{1}{2} {\cal A}_3 \right) \right],\nn\\
{\cal{R}}_2(\tau)&=&-\left(\Dot{{\cal P}} +\frac{{\cal P}}{\tau}\right)\nn
\eeq
 with ${\cal A}_1$, ${\cal A}_2$ and ${\cal A}_3$ given by, 
${\cal A}_1 = \cosh{\xi} \LR \nU -  {\cal B}_{(0)} \RR $,
${\cal A}_2 = \cosh{\xi} \LR {\cal A}_{(0)} - 3{\cal B}_{(0)} \RR  \label{A2}$, and  
${\cal A}_3 =  \cosh{\xi}\, {\cal B}_{(0)}$. 

From Eq.~(\ref{eq:C-coeff}), it can be noticed that for Bjorken flow all the $C$ coefficients evolve independently. Moreover,  the coefficients  ${C}_{\kappa X}$ and ${C}_{\kappa Y}$ (also ${C}_{\omega X}$ and ${C}_{\omega Y}$) obey the same differential equations. This is the consequence of rotational symmetry in the transverse directions. 

Similarly, the boost-invariant form of the conservation law's for charge current and energy-momentum can be written as 
\beq
\dot{n}+\frac{n}{\tau}=0.\lab{eq:charge},\\ 
\dot{\varepsilon}+\frac{(\varepsilon+P)}{\tau}=0.\lab{eq:en}
\eeq
The set of Eqs. \rfn{eq:charge}, \rfn{eq:en} and \rfn{eq:C-coeff} can be solved numerically. The procedure is to first solve the Eqs.~\rfn{eq:charge} and \rfn{eq:en} to obtain the temperature $T$ and chemical potential $\mu$ as a function of proper time $\tau$. Once $T$ and $\nu$ are known we can determine proper time dependence of functions ${\cal L}$, ${\cal P}$, ${\cal R}_1$,${\cal R}_2$, ${\cal Q}_1$, ${\cal Q}_2$ appearing the evolution equation (\ref{eq:C-coeff}) and finally the $C$-coefficients.  
%
\subsection{Spin polarization observable}
\label{subsec:5.2}
Space time evolution of $C$-coefficients can be used to ascertain the spin polarization of particles at freeze-out. The spin polarization of particles is given by the average Pauli-Luba\'nski (PL) vector $\langle\pi^\star_{\mu}(p)\rangle$ in the rest frame of the particles. 
The average PL vector $\langle\pi_{\mu}(p)\rangle$ of particles with momentum $p$ emitted from a given freeze-out hypersurface is provided by the ratio~\cite{Florkowski:2018ahw}
\beq
\langle\pi_{\mu}\rangle=\frac{E_p\frac{d\Pi _{\mu }(p)}{d^3 p}}{E_p\frac{d{\cal{N}}(p)}{d^3 p}}.
\eeq
where $E_p\frac{d\Pi _{\mu }(p)}{d^3 p}$ is the total value of PL vector of particles with momentum $p$ and ${E_p\frac{d{\cal{N}}(p)}{d^3 p}}$ is the momentum density of all particles given in terms of the following integrals
\beq
E_p\frac{d\Pi _{\mu }(p)}{d^3 p} &=& -\f{ \cosh(\xi)}{(2 \pi )^3 m}
\int
\Delta \Sigma _{\lambda } p^{\lambda } \,
e^{-\beta \cdot p} \,
\tilde{\omega }_{\mu \beta }p^{\beta }. \label{PDPLV}\\
E_p\frac{d{\cal{N}}(p)}{d^3 p}&=&
\f{4 \cosh(\xi)}{(2 \pi )^3}
\int
\Delta \Sigma _{\lambda } p^{\lambda } 
\,
e^{-\beta \cdot p} \,. \label{eq:MD}
\eeq
Here, $\Delta \Sigma _{\lambda}$ is an element of freeze-out hypersurface. In the above Eqs. (\ref{PDPLV}) and (\ref{eq:MD}), integration over freeze-out hypersurface can be carried out very easily by parametrizing particle four momentum $p^{\lambda}$ in terms of rapidity $y_{p}$ and transverse mass $m_{T}$ as; $ p^{\lambda}=(m_T \cosh(y_p),p_x,p_y,m_T \sinh(y_p))$ and assuming that freeze-out takes at a constant value of proper time ($\Delta \Sigma_{\lambda}=u_{\lambda}{dx}{dy}{\tau d\eta}$). Finally, after performing the canonical boost we can obtain the following result for the polarization vector   
\beq
\langle\pi^{\star}_{\mu}\rangle&=&-\frac{1}{8m }\left[\begin{array}{c}
0 \\ \\ \\
\left(\frac{\sh(y_p) p_x}{m_T \ch(y_p)+m}\right)\left[\CHI\left(C_{\kappa X} p_y-C_{\kappa Y} p_x\right)+2 C_{\omega Z} m_T  \right]\\ +
  \frac{ \CHI \,p_x \ch(y_p)  \left(C_{\omega X} p_x+C_{\omega Y} p_y\right)}{m_T \ch(y_p)+m}\!+\!2 C_{\kappa Z} p_y  \!-\!\CHI C_{\omega X}{m}_T \\ \\ \\
\left(\frac{\sh(y_p) p_y}{m_T \ch(y_p)+m}\right)\left[\CHI\left(C_{\kappa X} p_y-C_{\kappa Y} p_x\right)+2 C_{\omega Z} m_T  \right] \\+ \frac{\CHI \,p_y \ch(y_p)  \left(C_{\omega X} p_x+C_{\omega Y} p_y\right)}{m_T \ch(y_p)+m}\!-\!2 C_{\kappa Z} p_x \!-\!\CHI C_{\omega Y}{m}_T \\ \\ \\
 -\left(\frac{m\ch(y_p)+m_T}{m_T \ch(y_p)+m}\right)\left[\CHI\left(C_{\kappa X} p_y-C_{\kappa Y} p_x\right)+2 C_{\omega Z} m_T  \right]
\\-\frac{\CHI \,m\,\sh(y_p) \left(C_{\omega X} p_x+C_{\omega Y} p_y\right)}{m_T \ch(y_p)+m} \\
\end{array}
\right] \nn\\ 
\lab{PLVPPLPRF}
\eeq
where  $\CHI\equiv\CHI\left( \hat{m}_T \right)=\left( K_{0}\left( \hat{m}_T \right)+K_{2}\left( \hat{m}_T \right)\right)/K_{1}\left( \hat{m}_T \right)$ with $\hat{m}_T=m_T/T$. We can see that the time component of the polarization vector $\langle\pi^{\star}_{\mu}\rangle$ vanishes, this is because, in the particle rest frame we must have $\langle\pi^{\star}_{\mu}\rangle p^{\mu}=\langle\pi^{\star}_{0}\rangle m =0$. 
\subsection{Transverse momentum dependence of spin polarization}
\label{subsec:5.3}
\begin{figure}[ht!]
\centering
\includegraphics[width=0.56\textwidth]{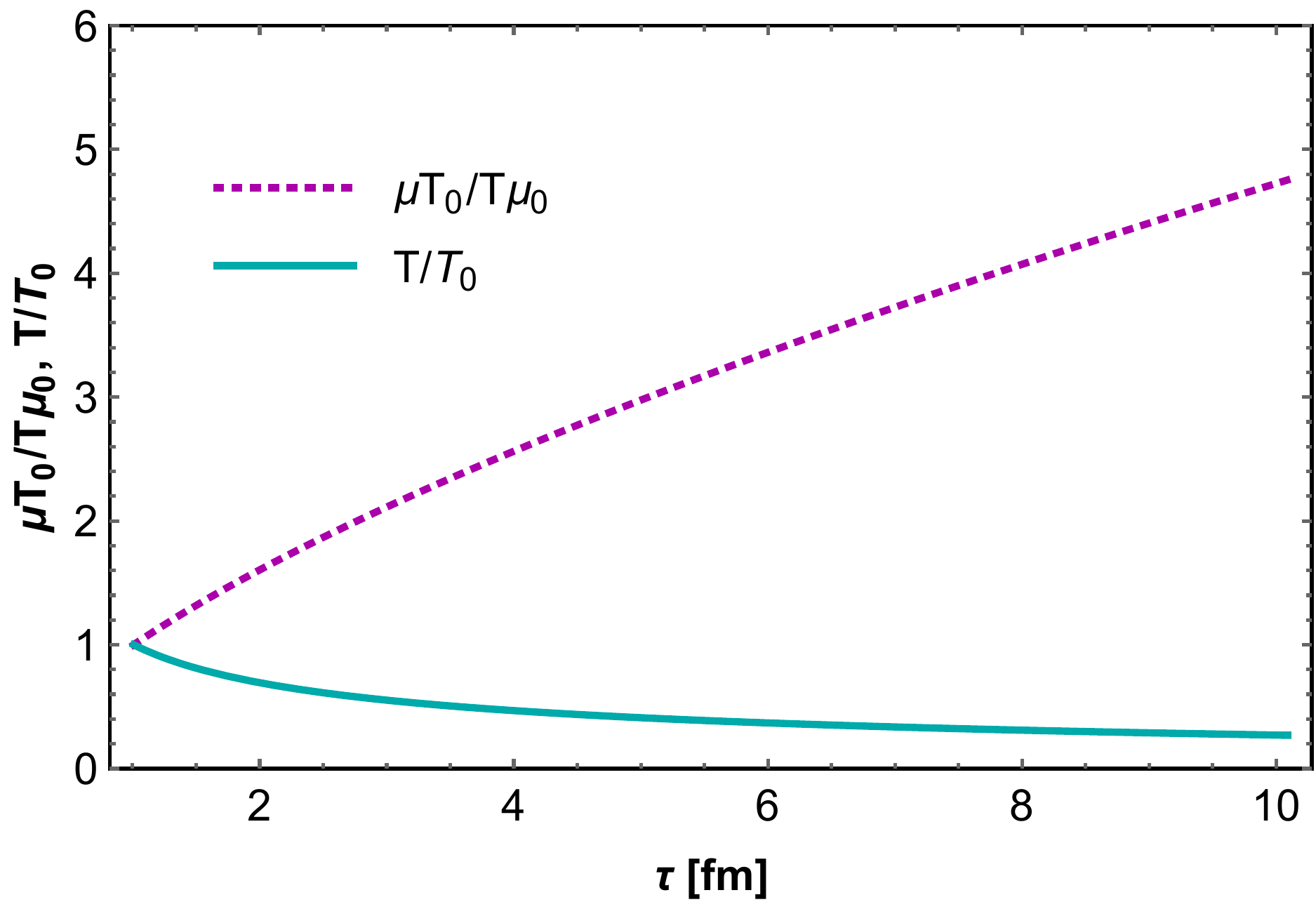}
\caption{Proper-time evolution of the ratio of baryon chemical potential to temperature ($\mu/T$) scaled to its initial value ($\mu_0/T_0$) and the temperature $T$ scaled to its initial value $T_0$}
\label{fig:t-mu}
\end{figure}

In the present section, we put forward numerical results for the spin polarization as a function of transverse momentum. We first solve the system of Eqs. (\ref{eq:charge}), (\ref{eq:en}) and (\ref{eq:C-coeff}) to obtain the dynamical evolution of $C$-coefficients. In order to study similar situation as in experiments we consider baryon rich matter with the initial baryon chemical potential $\mu_0=800$~MeV and the initial temperature $T_0=155$~MeV and assume that the system is composed of $\Lambda$-particles with their mass  $m~=~1116$~MeV. We continue the hydrodynamic evolution from initial proper time $\tau_0=1$~fm, till the final time $\tau_f=$~10~fm. 
By solving Eqs. (\ref{eq:charge}) and (\ref{eq:en}), in Fig. (\ref{fig:t-mu}) we show the proper-time dependence of the
temperature $T$ scaled by its initial temperature $T_0$ and the ratio ($\mu/T$) of the baryon chemical potential and temperature scaled by its initial value ($\mu_0/T_0$). It can be noticed that here we have reproduced a known result that $T/T_0$ decreases with increasing proper time, while the ratio of the chemical potential and the temperature to its initial values increases. 

\begin{figure}[ht!]
\centering
\includegraphics[width=0.56\textwidth]{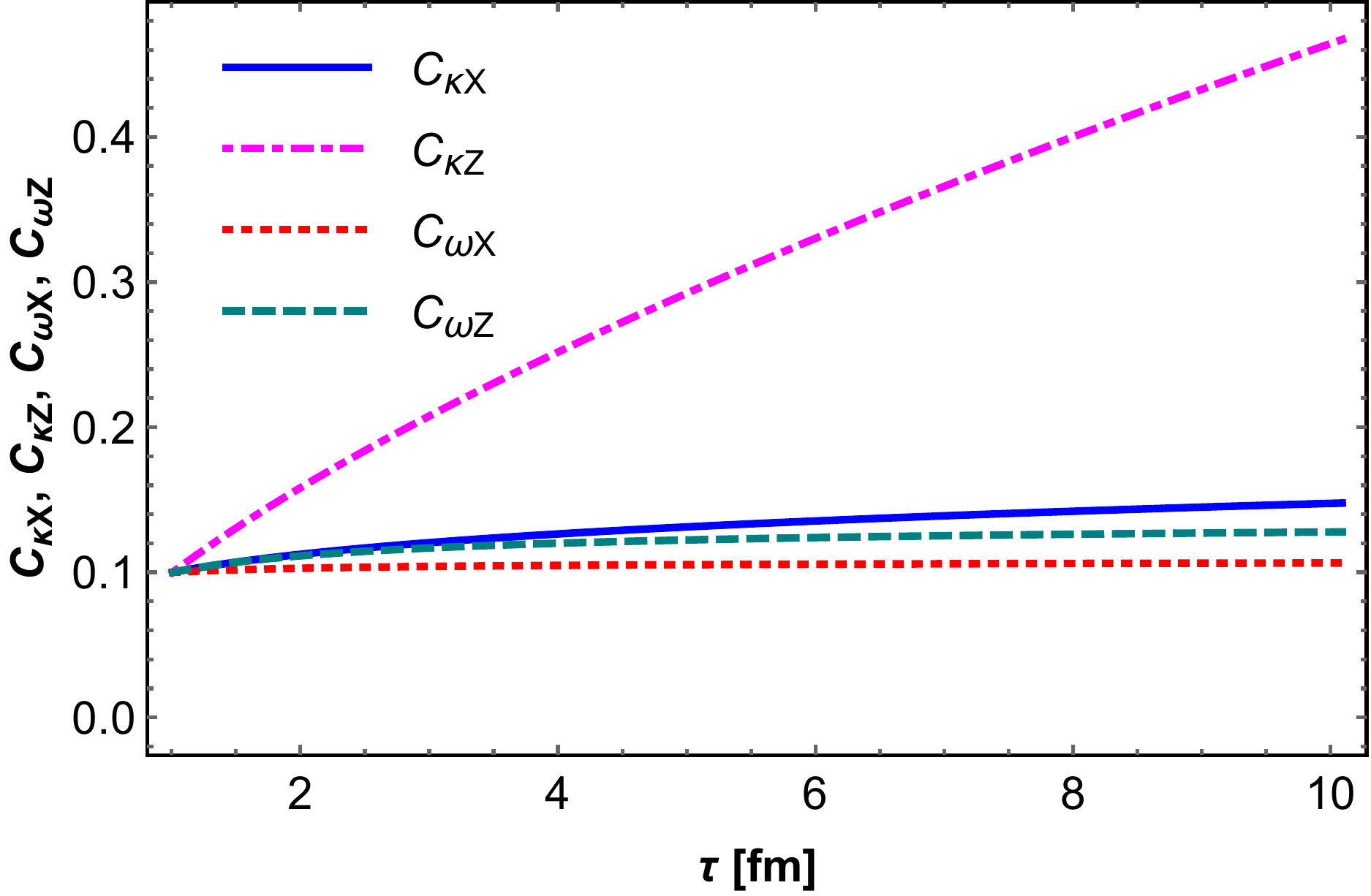}
\caption{Evolution of the coefficients $C_{\kappa X}$, $C_{\kappa Z}$, $C_{\omega X}$ and $C_{\omega Z}$ as a function of proper time. }
\label{fig:c_coef}
\end{figure}

Proper time evolution of various $C$-coefficients is determined by solving Eqs. (\ref{eq:C-coeff}) where $T(\tau)$ and $\mu(\tau)$ obtained by solving Eqs. (\ref{eq:charge}) and (\ref{eq:en}) used for background evolution. 
In Fig. (\ref{fig:c_coef}) we show the proper-time evolution of coefficients $C_{\kappa X}$, $C_{\kappa Z}$, $C_{\omega X}$ and $C_{\omega Z}$. In order to compare their relative dependence on proper time  we choose the same initial values (0.1) of all the $C$ coefficients. As we mentioned earlier that due to rotational symmetry in the transverse plane 
 $C_{\kappa Y}$ and $C_{\omega Y}$ fulfill the same equations as $C_{\kappa X}$ and $C_{\omega X}$, we have not shown their evolution in Fig. (\ref{fig:c_coef}). One can observe that the coefficient $C_{\kappa Z}$ has the strongest proper-time dependence as it increases by about 0.1 within 1 fm. 

\begin{figure}[ht!]
\centering
\subfigure[]{}\includegraphics[width=0.46\textwidth]{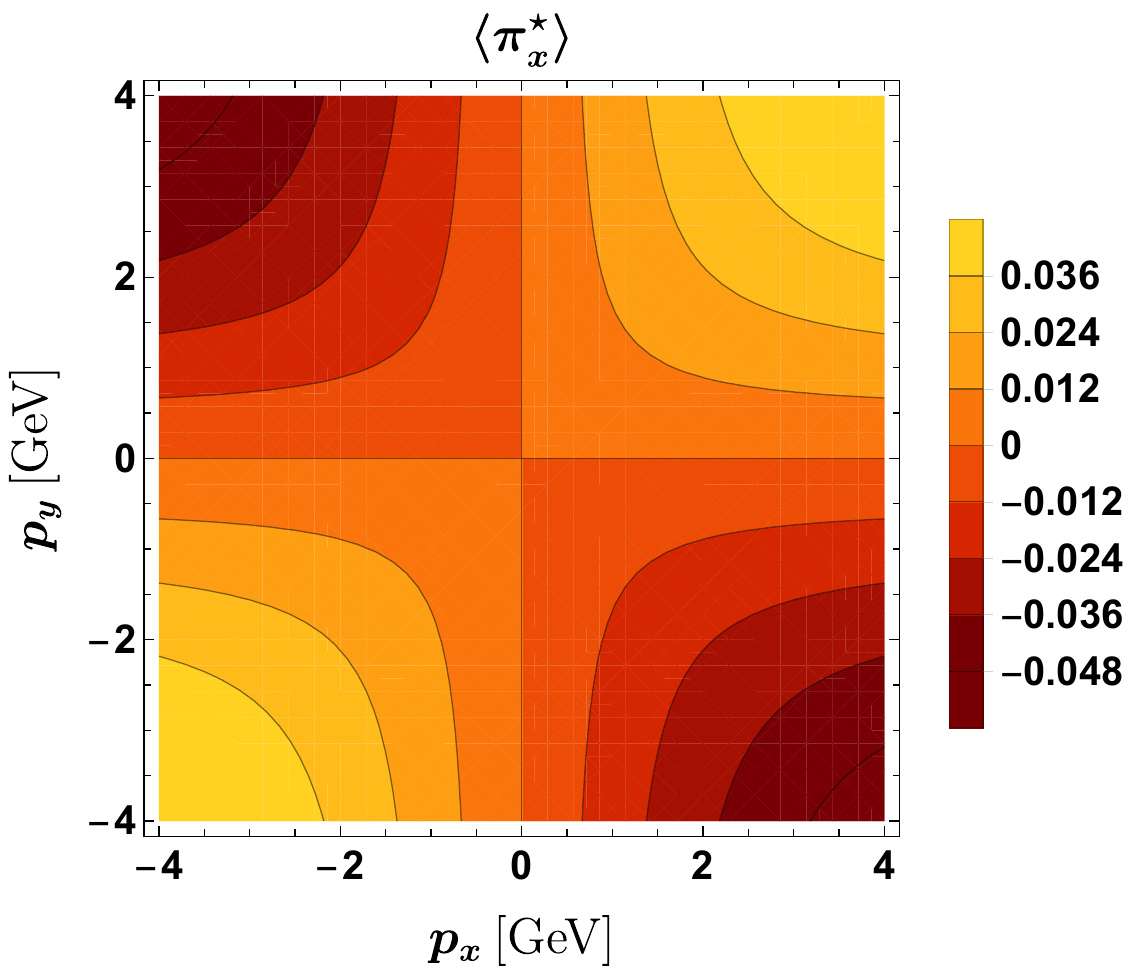}
\subfigure[]{}\includegraphics[width=0.46\textwidth]{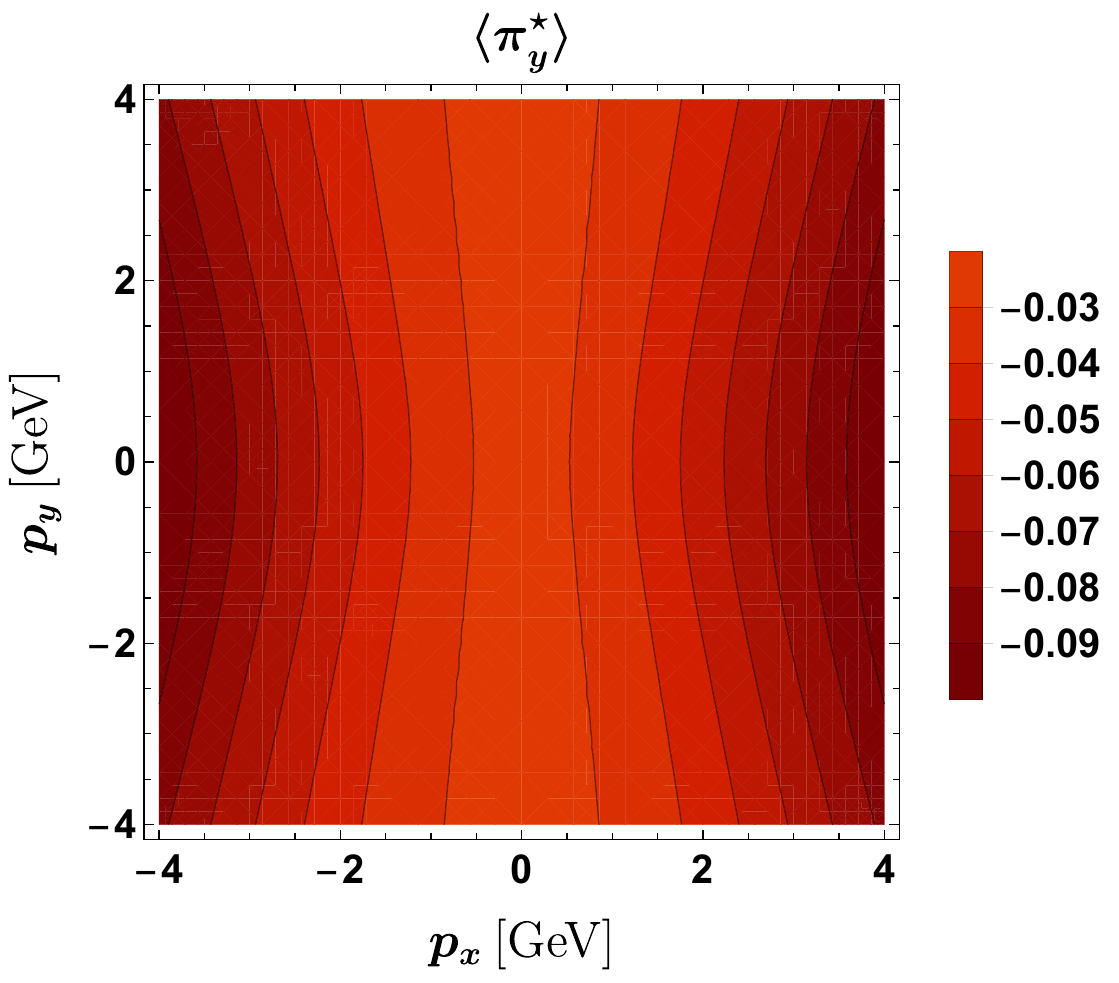}\\
\subfigure[]{}\includegraphics[width=0.46\textwidth]{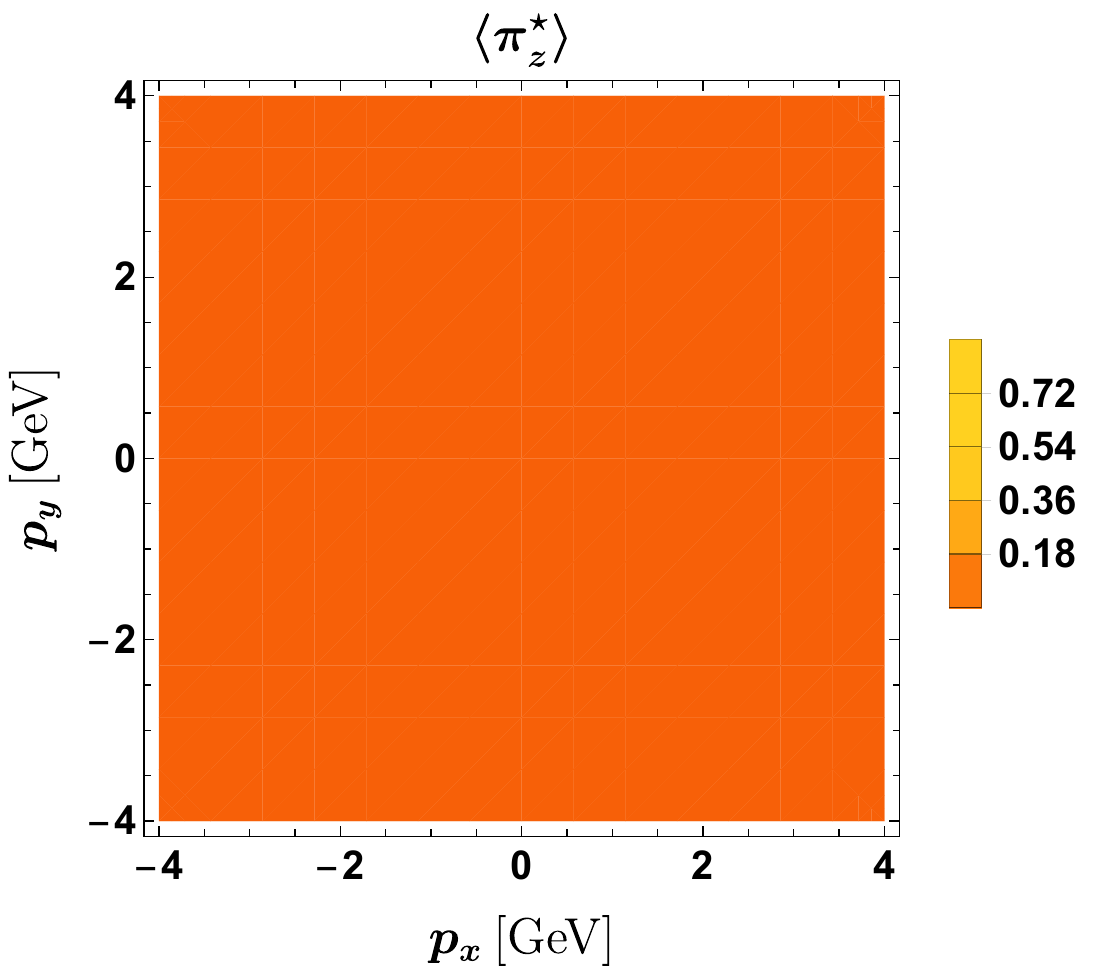}
\caption{Various components of the particle rest frame mean polarization vector of $\Lambda$'s as a function of transverse momentum obtained using the initial conditions $\mu_0=800$~MeV,  $T_0=155$~MeV, $\Cv_{\kappa, 0} = (0,0,0)$, and $\Cv_{\omega, 0} = (0,0.1,0)$ at  $y_p=0$.} 
\label{fig:polarization1}
\end{figure}

Knowing the proper time evolution of $C$-coefficients, we can determine the components of the mean polarization vector in the particle rest frame $\langle\pi^{\star}_{\mu}\rangle$ at freeze-out as functions of particle's transverse momentum. In Fig.~(\ref{fig:polarization1}) we show the numerical results for the different components of $\langle\pi^{\star}_{\mu}\rangle$ at mid particle rapidity {\it i.e.} $y_p=0$ for $\Lambda$-particles using initial conditions $\mu_0=800$~MeV, $T_0=155$~MeV, $\Cv_{\kappa, 0} = (0,0,0)$, and $\Cv_{\omega, 0} = (0,0.1,0)$. Note that the choice of the initial conditions $\Cv_{\kappa, 0} = (0,0,0)$, and $\Cv_{\omega, 0} = (0,0.1,0)$  of $C$-coefficients was made to address the physical situation that the initial spin angular momentum and total angular momentum of the colliding system are along the same direction. We observe that the component $\langle\pi^{\star}_{y}\rangle$ is negative, reflecting our choice of direction of spin angular momentum of the system. Due to the assumption $y_p=0$, the longitudinal component $\langle\pi^{\star}_{z}\rangle$ is zero which does not agree with observed quadrupole structure in experiments.  The reason for disagreement with experimental result is that 
quadrupole structure of $\langle\pi^{\star}_{z}\rangle$ appears in connection with the inhomogeneities in the transverse plane and the formation of the elliptic flow. Clearly, Bjorken symmetry does not offer this. In the case of $\langle\pi^{\star}_{x}\rangle$ we observe a quadrupole structure with changing signs in subsequent quadrants. Interestingly, the signs in the subsequent quadrants observed here are  opposite to the one obtained in other hydrodynamical calculations in Refs. \cite{Karpenko:2016jyx}. 
\subsection{Thermal Evolution of rotational fluid created in heavy ion collisions}
The presence of the spin-vorticity coupling in equilibrium distribution function can possibly modify the thermodynamic relation \cite{Becattini:2009wh,Florkowski:2017ruc},
 \begin{equation}
\varepsilon+P=Ts+\mu n+\Omega w,
\label{eos1}
\end{equation}
where, $\varepsilon$, $P$, $T$, $s$, $\mu$, $n$ respectively represent energy density, pressure, temperature, entropy density, chemical potential, number density. In the last term on the right hand side of Eq.~\eqref{eos1}, $\Omega$ is the vorticity  defined as $\Omega=\frac{T}{2\sqrt{2}} \sqrt{\omega^{\mu \nu}\omega_{\mu \nu}}$ where $\omega_{\mu \nu}$ is the spin polarization tensor. One may also regard $\Omega$ as the spin chemical potential which corresponds to spin density $w$. For a system in thermodynamic equilibrium, $\Omega$ is proportional to thermal vorticity~\cite{Becattini:2013fla}. This relation can influence the thermal evolution of the rotation fluid created in the relativistic heavy-ion collision~\cite{Bhatt:2018xsx}.  One needs to examine Eq.~\eqref{eos1} carefully when a local thermodynamic equilibrium is considered
~\cite{Becattini:2018duy}. It should be noted that Eq.~\eqref{eos1} is obtained for a phenomenological spin tensor ~\cite{Florkowski:2017ruc,Florkowski:2018fap} which is conserved.
However, it was proven that different choices of the energy-momentum and spin tensors are connected through pseudo gauge transformations~\cite{Florkowski:2018fap}. Here, for the purpose of  studying the thermal evolution, the spin degree of freedom has been considered to be fully equilibrated. This makes  spin as a hydrodynamical variable not important in our analysis. 
In this situation Eq.~\eqref{eos1} shall be invariant under the pseudo-gauge transformation, but it still has contribution from spin-orbit coupling.
The use of Eq.~\eqref{eos1} may provide some useful insight about the influence of spin-polarization on the thermal evolution of the fire ball. 
Further, we neglect the effect of viscosity by restricting ourselves to the regime of large Reynolds numbers.
 
Next, we need to consider the effect of finite vorticity on the velocity profile by doing similar to given in Ref.~\cite{Ollitrault:2007du}. First consider 4-velocity $u^{\mu}=(\gamma,\gamma v_x,0,\gamma v_z)$ where $\gamma $ is the Lorentz factor. Here fluid velocity satisfy the  condition $u^{\mu}u_{\mu}=1$. The vorticity is considered to be  along $\hat{y}$ direction and thus  $v_y=0$.  $v_z$ and $v_x$ are respectively the longitudinal and the transverse components of velocity.  In a manner similar to Ref.~\cite{Deng:2016gyh}, we assume the fluid velocity $\vec{v}$ is decomposed into two parts; (i) non-rotational flow ($\vec{v}_0$) and (ii) the rotational flow ($\vec{v}_r$).
The rotation flow velocity is defined by~\cite{Bhatt:2018xsx},
\beq
\vec{v}_r&=&\frac{1}{2}\vec{\omega}\times\vec{r} \label{eq:rof}
\eeq
Using the formula (\ref{eq:rof}), along with longitudinal expansion $v_x$ and $v_z$ can be written as
\beq
v_x=\frac{1}{2} \omega z, \quad \quad 
v_z=\frac{z}{\tau}-\frac{1}{2}\omega x. \label{109}
\eeq
In Eq.~\eqref{109} $x$ and $z$ are the positions coordinates, $v_x$ is due to the vorticity which vanishes in the case of non-vortical fluids. Similarly, for $\omega=0,$ the velocity, $v_z$ coincides with the velocity in the Bjorken flow. 

The equation for temperature evolution can be found from the projection of equation $\partial_{\mu}T^{\mu \nu}=0$ along the fluid four-velocity $u_\nu$ as,
\begin{equation}
 u_{\nu}\partial_{\mu}T^{\mu \nu}=0,
\end{equation}
which leads to
\begin{equation}
\partial_{\mu}((\varepsilon+P)u^{\mu})=u_{\nu}g^{\mu \nu}\partial_{\mu}P.
\label{evol}
\end{equation}
%

\begin{figure}[ht!]
\centering
\includegraphics[width=6.8cm]{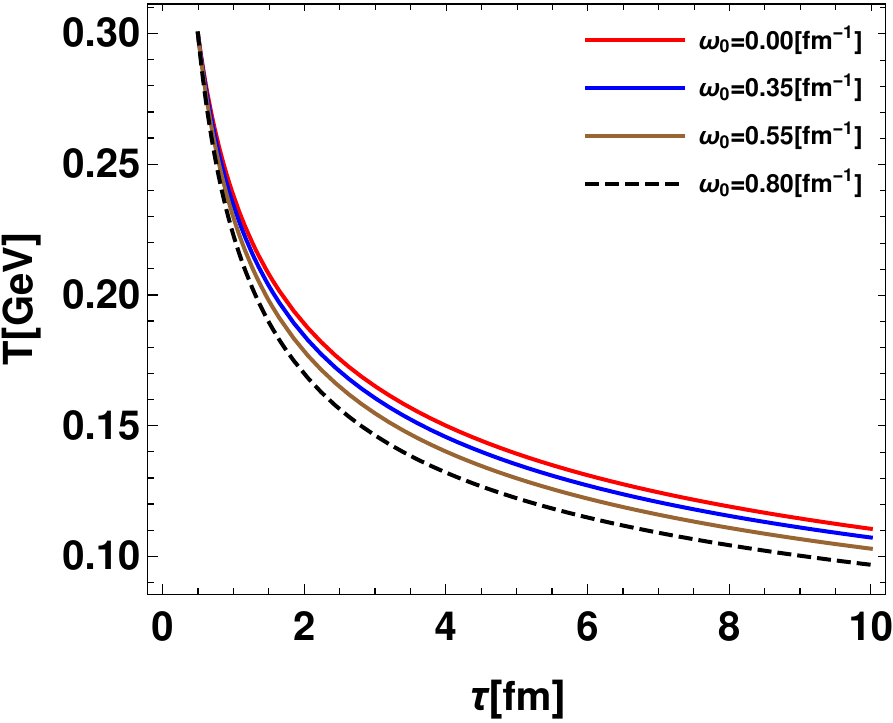}
\caption{
 Temperature(T) vs. $\tau$ plots for different values of the rotational parameter $\omega_0$ with initial time  $\tau_{0}=0.5$ fm and $T_{0}=300$ MeV. Red curve depicts the usual 1-D Bjorken flow case.} 
\label{flow}
\end{figure}

Now if one uses the modified thermodynamic relation Eq.~\eqref{eos1} for $\mu=0$ with spin-vorticity coupling together with $\varepsilon=3P$ \& $c_s^2=\frac{d \varepsilon}{d P}$,  
one now gets the following equation for the temperature evolution:
\begin{equation}
\frac{dT}{d\tau}=-c_s^2 \bigg(T+\frac{\omega w}{2 s}\bigg) (\partial_{\mu} u^{\mu}).
\label{evolution1}
\end{equation} 
Using the velocity profile defined above, one can numerically solve the temperature evolution equation. For a non-central collision with impact parameter of $b=7$ fm and the rms widths are $\sigma_x=2$ fm, $\sigma_y=2.6$ fm, $c_s^2=\frac{1}{3}$, $\tau_0=0.5$ fm and $T_0=300$ MeV, plots of temperature vs. time are shown for different values of vorticity.
The vorticity profile as given by
\begin{equation}
\vec{\omega}(\vec{r},t)=\frac{\omega_{0}{(\vec{r}_0,t_0)} t_0}{t}e^{-\frac{c_s^2}{2 \sigma_y^2}(t^2-t_{0}^2)}\hat{y},
\label{vorticity}
\end{equation}
in Eq.~\eqref{vorticity}, where $\omega_0$ is a free parameter; for details, see Ref.~\cite{Bhatt:2018xsx}. 
There is a one constraint on rotational motion that it required to satisfy the condition $\omega R < 1$.
Fig.~\ref{flow} shows the plot of temperature ($T$) vs time($\tau$). The red curve represents the 1D Bjorken flow which is a special case in our analysis when the vorticity is set to be zero.  As it can be gleaned from Fig.~\ref{flow}, coupling of the spin-vorticity leads to a faster cooling of the expanding fireball. This also leads to a reduction of the hadronization time as can be seen from table~\ref{table1}.
Note that we considered the initial temperature $T_0=300$ MeV and have taken critical temperature $T_c=150$ MeV. It can be noticed that with the increase in vorticity, the critical time (time scale for the system to reach critical temperature) decreases.

\begin{table}[ht!]
\centering
\begin{tabular}{|c|c|c|c|}
\hline
 $T_0$(MeV) & $T_c$ (MeV)&$\omega_0$ (fm$^{-1}$)&$\tau_c$ (fm) \\ 
\hline
&  & 0.0 & 4.05   \\
&  & 0.2  & 3.90   \\
&  & 0.4  & 3.79   \\
300.0& 150.0 & 0.6  & 3.48  \\
&  & 0.8  &  3.10 \\
&  & 0.9  &  2.75 \\

\hline
\end{tabular}
\caption{Critical time $\tau_c$ for different values of $\omega_0$ with $\tau_{0}=0.5$ fm.}
\label{table1}
\end{table}
\section{Summary}
\label{sec:6}
We have discussed the recent progress made in understanding the formulation of the framework of relativistic fluid dynamics with spin. 
Starting with the equilibrium Wigner functions with spin-1/2 particle and their semiclassical expansion, the fluid equations for the spin-polarized medium were obtained using the appropriate moments. By using classical treatment of spin-1/2 particles and the collisional invariance of the Boltzmann equation, the classical equilibrium distribution function for particles and antiparticle were introduced which depends on the spin polarization tensor. This tensor plays a role analogous to the chemical potential conjugate to the spin angular momentum. The equation of relativistic hydrodynamics for spin-polarized medium using these distribution functions is shown to be equivalent to the fluid equations obtained from the Wigner function formalism. We have also discussed very recent results of our paper~\cite{Bhadury:2020cop} where the effects of dissipation were included in the hydrodynamics. In this approach, a new set of kinetic coefficients associated with the spin are introduced. Next, we discussed the two applications of these ideal fluid equations to the relativistic heavy collision experiments. For the case transversely homogeneous longitudinal  expansion our results show, the spin polarization tensor can play a non-trivial role in the spin polarization of particles. In the presence of finite vorticity and the modification in the thermodynamic relation due to the vorticity can influence the thermal evolution of the early stages of the heavy-ion collisions. Here we would like to note that these applications are preliminary and they still require to be generalized to the more realistic cases.  Relativistic hydrodynamics for a spin-polarized medium is still under the early stages of developments and its applications to various realistic astrophysical and laboratory environments are yet to be explored.

\subsection*{Acknowledgment}
A.K. acknowledges the hospitality of National Institute of Science Education and Research where this work was done. A.K. was supported in part by the Department of Science and Technology, Government of India under the SERB NPDF Reference No. PDF/2020/000648. A.J. was supported in part by  the  DST-INSPIRE  faculty  award  under  Grant  No. DST/INSPIRE/04/2017/000038.

\appendix

\section{List of integrals in spin space}
\label{A1}
Here we list various formula used to carry out integration in spin space. The detail calculations will be presented in our paper~\cite{Bhadury:2020cop} 
\beq
\int \mathrm{dS} &=&\frac{m}{\pi \mathfrak{s}} \int \mathrm{d^4s}\, \delta(s\cdot s + {\mathfrak{s}}^2) \delta(p\cdot s)=2\nn\\
 \int\mathrm{dS}\,s^{\mu\nu} &=& 0. \nn\\
  \int \mathrm{dS}\, s_\sigma\, s_\delta &=& -\frac{2 \mathfrak{s}^2}{3} \bigg(g_{\sigma\delta} - \frac{p_\sigma p_\delta}{m^2}\bigg)\nn\\
   \int \mathrm{dS} s^{\mu\nu} s^{\alpha\beta} &= &- \frac{2 \mathfrak{s}^2}{3m^2} \epsilon^{\mu\nu\rho\sigma}\, \epsilon^{\alpha\beta\gamma\delta}\, p_\rho\, p_\gamma \bigg(g_{\sigma\delta} - \frac{p_\sigma p_\delta}{m^2}\bigg)
\eeq

\section{List of thermodynamic integrals $I_{nq}^{(r)}$}
\label{sec:thermint}

Thermodynamic integrals $I_{nq}^{(r)}$ are obtained by
\beq
I_{nq}^{(r)} &=& \frac{1}{(2q+1)!!} \int {dP}\,(u\cdot p)^{n-2q-r} (\Delta_{\alpha\beta} p^{\alpha} p^{\beta})^q e^{-\beta \cdot p}.
\eeq
From the above formula we can get,
\beq
I_{10}^{(0)} &=& \frac{T^3 z^2 }{2 \pi ^2} K_2(z)\nn\\
I_{20}^{(0)} &=& \frac{T^4 z^2}{2 \pi^2} \left[3 K_2(z) + zK_1(z)\right]\nn\\
I_{21}^{(0)} &=& - \frac{T^4 z^2}{2 \pi ^2} K_2(z)\nn\\
I_{30}^{(0)} &=& \frac{z^5 T^5}{2\pi^2} \left[K_5(z)+K_3(z)-2K_1(z)\right]\nn\\
I_{31}^{(0)} &=& - \frac{z^5T^5}{96\pi^2} \left[K_5(z)-3K_3(z)+2K_1(z)\right]\nn\\
I_{40}^{(0)} &=& \frac{T^6 z^6}{64 \pi^2} \left[K_{6}(z) + 2K_{4}(z) - K_{2}(z) - 2K_{0}(z)\right]\nn\\
I_{41}^{(0)} &=& - \frac{T^6 z^6}{192 \pi^2} \left[K_{6}(z) - 2K_{4}(z) - K_{2}(z) + 2K_{0}(z)\right]\nn\\
I_{42}^{(0)} &=& \frac{T^6 z^6}{960 \pi^2} \left[K_{6}(z) - 6K_{4}(z) + 15K_{2}(z) - 10K_{0}(z)\right]\nn\\
I_{21}^{(1)} &=& - \frac{T^3 z^3}{6 \pi^2} \left[\frac{1}{4} K_3(z) - \frac{5}{4} K_1(z) + K_{i,1}(z)\right]\nn\\
I_{42}^{(1)} &=& \frac{T^5 z^5}{480 \pi^2} \Big[22K_{1}(z) - 7K_{3}(z) + K_5(z) - 16K_{i,1}(z)\Big]\nn
\eeq
In the above formulas, $K_n(z)$  are the modified Bessel functions of the second kind while $K_{i,1}(z)$ are the first order Bickley-Naylor function with the argument $z = m/T$. The function $K_n(z)$ and $K_{i,1}(z)$ are expressed as
\beq
K_n(z) &=& \int_0^{\infty} dx\, \cosh{nx}\, e^{- z \cosh x}.\\
K_{i,1}(z) &=& \int_0^{\infty } dx \sech{x}\, e^{-z \cosh x}\nn\\
&=& \frac{\pi}{2} \Big[ 1 - z\, K_{0}(z)\, L_{-1}(z) - z\, K_{1}(z)\, L_{0}(z)\Big]
\eeq
In the expression for $K_{i,1}(z)$, function $L_i$ is the modified Struve function.

Note that here we have not listed the function $I_{20}^{(1)}$, $I_{30}^{(1)}$, $I_{31}^{(1)}$,  $I_{40}^{(1)}$, $I_{50}^{(1)}$, $I_{51}^{(1)}$ and $I_{52}^{(1)}$ as they all can be written in terms of above listed integrals using the recurrence relation as given below. 
\bea
I_{\rm n,q}^{(r)} &=& I_{\rm n-1,q}^{(r-1)};~~~ n\geq2 q, \label{eq:ir1}\\
I_{\rm n,q}^{(0)} &=& \frac{1}{\beta} \left[(n - 2 q) I_{\rm n-1,q}^{(0)} - I_{\rm  n-1, q-1}^{(0)}\right],\label{eq:ir2}\\
\dot{I}_{\rm n,q}^{(0)} &=& - \dot{\beta} I_{n+1, q}^{(0)}\label{eq:ir3}
\eea  

\section{List of D-coefficients}
\label{AD3}

Expressions for various D-coefficients are as follows
\beq
D_{\Pi}^{\mu \nu} &=& D_{\Pi 1} \omega^{\mu \nu} + D_{\Pi 2} u^{\alpha} u^{[\mu} \omega^{\nu]}{}_{\alpha}\label{eq:DPi1}\\
D_n^{[\mu \nu]}{}_{\alpha} &=& - D_{n1}\left(u^{[\mu} \omega^{\nu]}{}_{\alpha} + g^{[\mu}{}_{\alpha} u^{\kappa} \omega^{\nu]}{}_{\kappa}\right) - D_{n2} u^{[\mu} \Delta^{\nu]}{}_{\rho} \omega^{\rho}{}_{\alpha}\label{eq:Dnmunua1.1}\\
D_{\pi}^{[\mu}{}_{\lambda} &=& - \omega^{[\mu}{}_{\lambda} \frac{4 I_{31}^{(0)}}{(m^2\, I_{10}^{(0)} - 2 I_{31}^{(0)})} - u^{[\mu} u^{\alpha} \omega_{\alpha \lambda} \frac{4 (I_{30}^{(0)} - I_{31}^{(0)}) I_{31}^{(0)}}{(m^2\, I_{10}^{(0)} - 2\, I_{31}^{(0)}) \big[m^2\, I_{10}^{(0)} - (I_{30}^{(0)} + I_{31}^{(0)})\big]}\label{eq:Dnulam1}\\
D_{\Sigma 1}^{\alpha} &=& - u^{\alpha} \frac{2\, I_{31}^{(0)}}{(m^2\, I_{10}^{(0)} - 2\, I_{31}^{(0)})} \label{eq:Dsig1.1}\\
D_{\Sigma 2}^{[\mu \nu] \alpha} &=& - u^{[\mu} g^{\nu]\alpha} \frac{2\, I_{31}^{(0)}}{ \left(m^2\, I_{10}^{(0)} - 2\, I_{31}^{(0)}\right)} - u^{[\mu} \Delta^{\nu]\alpha} \frac{2\, (I_{30}^{(0)} - I_{31}^{(0)}) I_{31}^{(0)}}{\left( m^2\, I_{10}^{(0)} - 2\, I_{31}^{(0)}\right) \left[m^2\, I_{10}^{(0)} - \left( I_{30}^{(0)} + I_{31}^{(0)}\right)\right]}\nn\\\label{eq:Dsig2.1}
\eeq
where
\beq
D_{\Pi 1} &=& - \frac{1}{\left(I_{10}^{(0)} - \frac{2}{m^2} I_{31}^{(0)}\right)} \left(\xi_{\theta} \tanh \xi\, I_{10}^{(0)} - \beta_{\theta} I_{20}^{(0)} + I_{10}^{(0)} - \frac{2}{m^2} \xi _{\theta}\, \tanh \xi\, I_{31}^{(0)} + \frac{2\, \beta_{\theta}\, I_{41}^{(0)}}{m^2} - \frac{10 I_{31}^{(0)}}{3\, m^2}\right)\label{eq:a1.1}\\
D_{\Pi 2} &=& \frac{2}{m^2\, I_{10}^{(0)} - 2\, I_{31}^{(0)}} \Bigg[\beta_{\theta} \!\left(I_{40}^{(0)} - I_{41}^{(0)}\right) - \xi_{\theta}\! \left(I_{30}^{(0)} - I_{31}^{(0)}\right) \tanh \xi - \left(I_{30}^{(0)} - \frac{11}{3} I_{31}^{(0)}\right) + \frac{\left(I_{30}^{(0)} - I_{31}^{(0)}\right)}{m^2\, I_{10}^{(0)} - I_{30}^{(0)} - I_{31}^{(0)}}\nn\\
&& \!\times\bigg(\!m^2\, \xi_{\theta} \tanh\xi\, I_{10}^{(0)} - m^2 \beta_{\theta}\, I_{20}^{(0)} + m^2 I_{10}^{(0)} - \xi_{\theta}\! \left(\!I_{30}^{(0)} + I_{31}^{(0)}\!\right)\! \tanh \xi + \beta_\theta\! \left(\!I_{40}^{(0)} + I_{41}^{(0)}\!\right)\! +\beta I_{41}^{(0)} - \frac{5}{3} I_{31}^{(0)}\bigg)\! \Bigg]\nn\\\label{eq:a2.1}\\
D_{n1} &=& \frac{2 \tanh \xi}{\left(m^2\, I_{10}^{(0)} - 2\, I_{31}^{(0)}\right)} \left(I_{31}^{(0)} - \frac{n_0\, I_{41}^{(0)}}{\varepsilon_0 + P_0}\right) \label{eq:R_1.1}\\
D_{n2} &=& \frac{\tanh \xi}{m^2I_{10}^{(0)} - \left(I_{30}^{(0)} + I_{31}^{(0)}\right)} \left(I_{31}^{(0)} - \frac{n_0 I_{41}^{(0)}}{\varepsilon_0 + P_0}\right) \frac{2 \left(I_{30}^{(0)} - I_{31}^{(0)}\right)}{\left(m^2\, I_{10}^{(0)} - 2\, I_{31}^{(0)}\right)} \label{eq:R_3.1}
\eeq

\section{List of C-coefficients}

Various C-coefficients are given by following expressions
\label{AC3}
\beq
C_{\Pi} &=& - \frac{1}{m^2 I_{10}^{(0)} - \left(I_{30}^{(0)} + I_{31}^{(0)}\right)} \Bigg[m^2 \xi_{\theta} \tanh \xi I_{10}^{(0)} - m^2\, \beta_\theta\, I_{20}^{(0)} + m^2\, I_{10}^{(0)} - \tanh \xi \left(I_{30}^{(0)} + I_{31}^{(0)}\right) \xi_{\theta} \nn\\
&& \qquad\qquad\qquad\qquad\qquad\qquad + \beta_{\theta} \left(I_{40}^{(0)} + I_{41}^{(0)}\right) + \beta\, I_{41}^{(0)} - \frac{5}{3} I_{31}^{(0)}\Bigg] \label{eq:CPI}\\
C_{n} &=& \frac{\tanh \xi}{m^2\, I_{10}^{(0)} - \left(I_{30}^{(0)} + I_{31}^{(0)}\right)} \left(I_{31}^{(0)} - \frac{n_0 I_{41}^{(0)}}{\varepsilon_0 + P_0}\right)\label{eq:CN}\\
C_{\pi} &=& - \frac{2 I_{31}^{(0)}}{m^2\, I_{10}^{(0)} - \left(I_{30}^{(0)} + I_{31}^{(0)}\right)} \label{eq:Cpi}\\
C_{\Sigma} &=& \frac{I_{31}^{(0)}}{m^2\, I_{10}^{(0)} - \left(I_{30}^{(0)} + I_{31}^{(0)}\right)}\label{CSIG}
\eeq

\section{List of $\beta$-coefficients}

\label{AB3}
\begin{eqnarray}
\beta_{\Pi}^{(1)} &=& \frac{4\, \mathfrak{s}^2}{3}\Bigg(\!-\frac{2}{m^2} \xi_{\theta}\, \sinh \xi\, I_{41}^{(1)} + \frac{2}{m^2} I_{51}^{(1)} \beta_{\theta}\, \cosh \xi + \frac{10}{3m^2} I_{52}^{(1)}  \beta\, \cosh \xi - \frac{2}{m^2} I_{41}^{(1)} \cosh \xi\, D_{\Pi 1}\Bigg)\label{eq:betaPi1}\\
\beta_{\Pi}^{(2)} &=& \frac{4\, \mathfrak{s}^2}{3} \Bigg[\!-\frac{2}{m^2} \xi_{\theta}\, \sinh \xi I_{40}^{(1)} + \frac{4}{m^2} \xi_{\theta}\, \sinh \xi\, I_{41}^{(1)} + \frac{2}{m^2} I_{50}^{(1)} \beta_{\theta}\, \cosh \xi + \frac{2}{m^2} I_{51}^{(1)} \beta\, \cosh \xi - \frac{4}{m^2} I_{51}^{(1)} \beta_\theta\, \cosh \xi \nn\\
&-& \frac{20}{3m^2} I_{52}^{(1)} \beta\, \cosh \xi - \left(I_{20}^{(1)} - \frac{3}{m^2} I_{41}^{(1)}\right) \cosh \xi\, D_{\Pi 2} - \frac{2}{m^2} \left(I_{40}^{(1)} - 2\, I_{41}^{(1)}\right) \cosh \xi\, C_\Pi\Bigg] \label{eq:betaPi2}\\
\beta_{\Pi }^{(3)} &=& \frac{4\, \mathfrak{s}^2}{3}\Bigg( - \frac{2}{m^2} \xi_{\theta}\, \sinh \xi I_{41}^{(1)} + \frac{2}{m^2} I_{51}^{(1)} \beta_{\theta}\, \cosh \xi + \frac{10}{3m^2} I_{52}^{(1)} \beta\, \cosh \xi - \frac{2}{m^2} I_{41}^{(1)} \cosh \xi\, C_\Pi\Bigg) \label{eq:betaPi3}
\end{eqnarray}
\begin{eqnarray}
\beta_{\pi}^{(1)} &=& \frac{16\, \mathfrak{s}^2}{3\, m^2} \beta\, \cosh \xi\, I_{42}^{(0)}
\label{eq:betapi1}\\
\beta_{\pi}^{(2)} &=&\frac{16\, \mathfrak{s}^2}{3\, m^2} \cosh \xi \left(\beta\, I_{42}^{(0)} - \frac{ I_{41}^{(1)}\, I_{31}^{(0)}}{m^2\, I_{10}^{(0)} - 2\, I_{31}^{(0)}}\right)
\label{eq:betapi2}\\
\beta_{\pi}^{(3)} &=& \frac{16\, \mathfrak{s}^2}{3\, m^2} \cosh \xi \left(\frac{I_{41}^{(1)}\, I_{31}^{(0)}}{m^2\, I_{10}^{(0)} - 2\, I_{31}^{(0)}}\right)
\label{eq:betapi3}\\
\beta_{\pi}^{(4)} &=& \frac{16\, \mathfrak{s}^2}{3\, m^2} \cosh \xi \left(\frac{I_{41}^{(1)}\, I_{31}^{(0)}}{m^2 I_{10}^{(0)}-\left(I^{(0)}_{30}+I_{31}^{(0)}\right)}\right)
\label{eq:betapi4}
\end{eqnarray}
\begin{eqnarray}
\beta_{n}^{(1)} &=& \frac{4 \mathfrak{s}^2}{3\, m^2} \cosh \xi \left[-\tanh \xi \left(m^2\, I_{21}^{(1)} - 2\, I_{42}^{(1)}\right) + \bigg(\frac{n_0 \tanh(\xi)}{\varepsilon_0 + P_0} \bigg) \left(m^2\, I_{31}^{(1)} - 2\, I_{52}^{(1)}\right)\right]\label{eq:betan1}\\
\beta_{n}^{(2)} &=& \frac{8 \mathfrak{s}^2}{3\, m^2} \cosh \xi  \Bigg[\!-\tanh \xi \!\left(I_{41}^{(1)} - I_{42}^{(1)}\right) + \bigg(\frac{n_0 \tanh \xi }{\varepsilon_0 + P_0} \bigg)\!\! \left(I_{51}^{(1)} - I_{52}^{(1)}\right) - \frac{I^{(1)}_{41}\, \tanh \xi}{\left(m^2\, I_{10}^{(0)} - 2\, I_{31}^{(0)}\right)}\! \left(\!I_{31}^{(0)} - \frac{n_0 I_{41}^{(0)}}{\varepsilon_0 + P_0} \!\right) \!\! \Bigg]~~~~~~\nn\\ \label{eq:betan2}\\
\beta_{n}^{(3)} &=& \frac{8 \mathfrak{s}^2}{3\, m^2} \cosh \xi \left[ - \tanh \xi\, I_{42}^{(1)} + \bigg(\frac{n_0 \tanh \xi}{\varepsilon_0 + P_0}\bigg) I_{52}^{(1)}\right]
\label{eq:betan3}\\
\beta_{n}^{(4)} &=& \frac{8 \mathfrak{s}^2}{3\, m^2} \cosh \xi \left[\frac{I^{(1)}_{41}\, \tanh \xi}{\left(m^2 I_{10}^{(0)}-2 I_{31}^{(0)}\right)}\left(I_{31}^{(0)}-\frac{n_0 I_{41}^{(0)}}{\varepsilon _0+P_0}\right)\right]
\label{eq:betan4}\\
\beta_{n}^{(5)} &=& \frac{8 \mathfrak{s}^2}{3\, m^2} \cosh \xi \Bigg[ - \tanh \xi\, I_{42}^{(1)} + \bigg(\frac{n_0 \tanh \xi}{\varepsilon_0 + P_0} \bigg) I_{52}^{(1)} - \frac{I_{41}^{(1)}\, \tanh \xi}{m^2\, I_{10}^{(0)} - \left(I_{30}^{(0)} + I_{31}^{(0)}\right)} \left(I_{31}^{(0)} - \frac{n_0 I_{41}^{(0)}}{\varepsilon_0 + P_0} \right)\Bigg] \label{eq:betan5}\\
\beta_{n}^{(6)} &=& \frac{8 \mathfrak{s}^2}{3\, m^2} \cosh \xi \Bigg[ \frac{I_{41}^{(1)}\, \tanh \xi}{m^2\, I_{10}^{(0)} - \left(I_{30}^{(0)} + I_{31}^{(0)}\right)} \left(I_{31}^{(0)} - \frac{n_0 I_{41}^{(0)}}{\varepsilon_0 + P_0} \right)\Bigg] \label{eq:betan6}
\end{eqnarray}
\begin{eqnarray}
\beta_{\Sigma}^{(1)} &=& - \frac{4 \mathfrak{s}^2}{3} \cosh \xi\, I_{21}^{(1)} \label{eq:betaSig1}\\
\beta_{\Sigma}^{(2)} &=& - \frac{8 \mathfrak{s}^2}{3\, m^2} \cosh \xi \left(I_{41}^{(1)} + \frac{I_{41}^{(1)}\, I_{31}^{(0)}}{m^2 I_{10}^{(0)}-2 I_{31}^{(0)}}\right) \label{eq:betaSig2}\\
\beta_{\Sigma}^{(3)} &=& - \frac{8 \mathfrak{s}^2}{3\, m^2} \cosh \xi\, I_{42}^{(1)} \label{eq:betaSig3}\\
\beta_{\Sigma}^{(4)} &=& - \frac{8 \mathfrak{s}^2}{3\, m^2} \cosh \xi \left( \frac{I_{41}^{(1)}\, I_{31}^{(0)}}{m^2I_{10}^{(0)} - \left(I_{30}^{(0)} + I_{31}^{(0)}\right)}\right)\label{eq:betaSig4}\\
B_{\Sigma}^{(5)} &=& \frac{8 \mathfrak{s}^2}{3\, m^2} \cosh \xi \left(\frac{I_{41}^{(1)}\, I_{31}^{(0)}}{m^2\, I_{10}^{(0)} - 2\, I_{31}^{(0)}}\right) \label{eq:betaSig5}
\end{eqnarray}

\bibliography{spin}
\bibliographystyle{utphys}

\end{document}